\documentclass[reprint, superscriptaddress, amsmath, amssymb, aps, prl, floatfix,]{revtex4-2}

\usepackage{braket}
\usepackage{printlen} 
\usepackage{amsthm}
\usepackage{isotope}
\usepackage[
    per-mode=symbol,
    retain-zero-uncertainty=true
    ]{siunitx}
\DeclareSIUnit{\gauss}{G}
\DeclareSIUnit{\phonons}{phonons}
\usepackage{braket}
\usepackage{color}
\usepackage{graphicx}
\graphicspath{{supp_mat_figures/}}
\usepackage{dcolumn}
\usepackage{bm}

\usepackage{hyperref}
\hypersetup{
pdftitle={Low-excitation transport and separation of high-mass-ratio mixed-species ion chains},
pdfauthor={F. Lancellotti, S. Welte, M. Simoni, C. Mordini, T. Behrle, B. de Neeve, M. Marinelli, V. Negnevitsky, J. P. Home},
pdfkeywords = {Ion trapping, Ion transport, QCCD architecture},
colorlinks=true, linkcolor=[RGB]{0,0,0}, citecolor=magenta, filecolor=[RGB]{0,0,0}, urlcolor=[RGB]{0,0,0}}

\usepackage{bm}
\usepackage{enumerate}
\usepackage{tikz}
\usepackage{dsfont}
\usetikzlibrary{quantikz}
\usepackage{lipsum} 
\usepackage{subcaption}
\usepackage{ragged2e}
\DeclareCaptionJustification{justified}{\justifying}
\usepackage{printlen}
\usepackage{etoolbox}
\makeatletter
\def\maketitle{
\@author@finish
\title@column\titleblock@produce
\suppressfloats[t]}
\makeatother
\newcounter{PRLsections}

\newcounter{PRLsubsections}[section]

\usepackage{xpatch}
\xpretocmd{\section}{\setcounter{PRLsubsections}{0}}{}{}

\usepackage{wasysym}
\usepackage[OT2,T1]{fontenc}
\DeclareSymbolFont{cyrletters}{OT2}{wncyr}{m}{n}
\DeclareMathSymbol{\Sha}{\mathalpha}{cyrletters}{"58}

\usepackage[innercaption]{sidecap}
\sidecaptionvpos{figure}{c}
\DeclareSymbolFont{yhlargesymbols}{OMX}{yhex}{m}{n} 
\DeclareMathAccent{\widehat}{\mathord}{yhlargesymbols}{"62}
\usepackage{multirow}
\usepackage{array}
\newcolumntype{L}[1]{>{\raggedright\let\newline\\\arraybackslash\hspace{0pt}}m{#1}}
\newcolumntype{C}[1]{>{\centering\let\newline\\\arraybackslash\hspace{0pt}}m{#1}}
\newcolumntype{R}[1]{>{\raggedleft\let\newline\\\arraybackslash\hspace{0pt}}m{#1}}

\newcommand{\be}{\begin{equation}} 
\newcommand{\ee}{\end{equation}}

\usepackage{xspace}
\newcommand{\beryllium}{$\isotope[9]{Be}^+$\xspace}
\newcommand{\calcium}{$\isotope[40]{Ca}^+$\xspace}

\usepackage[normalem]{ulem}

\newtoggle{arXiv} 

\toggletrue{arXiv}      
\begin{document}
\preprint{APS/123-QED}

\title{Low-excitation transport and separation of high-mass-ratio mixed-species ion chains}

\author{F. Lancellotti}
\email[E-mail: ]{flancell@phys.ethz.ch}
\affiliation{Institute for Quantum Electronics, ETH Z\"urich, Otto-Stern-Weg 1, 8093 Z\"urich, Switzerland}
\author{S. Welte}
\email[E-mail: ]{weltes@phys.ethz.ch}
\affiliation{Institute for Quantum Electronics, ETH Z\"urich, Otto-Stern-Weg 1, 8093 Z\"urich, Switzerland}
\author{M. Simoni}
\affiliation{Institute for Quantum Electronics, ETH Z\"urich, Otto-Stern-Weg 1, 8093 Z\"urich, Switzerland}
\author{C. Mordini}
\affiliation{Institute for Quantum Electronics, ETH Z\"urich, Otto-Stern-Weg 1, 8093 Z\"urich, Switzerland}
\author{T. Behrle}
\affiliation{Institute for Quantum Electronics, ETH Z\"urich, Otto-Stern-Weg 1, 8093 Z\"urich, Switzerland}
\author{B. de Neeve}
\affiliation{Institute for Quantum Electronics, ETH Z\"urich, Otto-Stern-Weg 1, 8093 Z\"urich, Switzerland}
\author{M. Marinelli}
\altaffiliation{Current address: JILA, Boulder, Colorado, USA}
\affiliation{Institute for Quantum Electronics, ETH Z\"urich, Otto-Stern-Weg 1, 8093 Z\"urich, Switzerland}
\author{V. Negnevitsky}
\altaffiliation{Current address: Oxford Ionics, Oxford, OX5 1PF, United Kingdom}
\affiliation{Institute for Quantum Electronics, ETH Z\"urich, Otto-Stern-Weg 1, 8093 Z\"urich, Switzerland}
\author{J. P. Home}
\email[E-mail: ]{jhome@phys.ethz.ch}
\affiliation{Institute for Quantum Electronics, ETH Z\"urich, Otto-Stern-Weg 1, 8093 Z\"urich, Switzerland}
\affiliation{Quantum Center, ETH Z{\"u}rich, 8093 Z{\"u}rich, Switzerland}

\begin{abstract}
\medskip \noindent
We demonstrate low-excitation transport and separation of two-ion crystals consisting of one \beryllium and one \calcium ion, with a high mass ratio of $4.4$. The full separation involves transport of the mixed-species chain, splitting each ion into separate potential wells, and then transport of each ion prior to detection. We find the high mass ratio makes the protocol sensitive to mode crossings between axial and radial modes, as well as to uncontrolled radial electric fields that induce mass-dependent twists of the ion chain, which initially gave excitations $\bar{n}{\gg}10$. By controlling these stages, we achieve excitation as low as $\bar{n}{=}\SI{1.40 \pm 0.08}{\phonons}$ for the calcium ion and $\bar{n}{=}\SI{1.44 \pm 0.09}{\phonons}$ for the beryllium ion. Separation and transport of mixed-species chains are key elements of the QCCD architecture, and may also be applicable to quantum-logic-based spectroscopy of exotic species.
\end{abstract}

\maketitle

Trapped ions are among the leading candidates for quantum computing, based on the high quality of control demonstrated and long coherence times \cite{clark2023, wang2021}. A primary challenge remains to scale to sizes suitable for large-scale error correction. The Quantum Charge-Coupled Device (QCCD) \cite{wineland1998,kielpinski2002} is a promising candidate architecture for this scaling. A QCCD chip employs multiple separate zones to perform specialized tasks like quantum logic operations, storage of quantum memories, and qubit detection and reset. Ion transport and separation are used to connect separate zones thus making them necessary ingredients to perform large-scale operations \cite{bruzewicz2019}. The QCCD architecture places high demands on the quality of shuttling and separation operations, which, if performed imperfectly, cause motional excitation, degrading the performance of subsequent quantum gates \cite{kaushal2020}. Such excitation can be mitigated by co-trapping a second ion species and using it to re-cool the ions \cite{haffner2008, schmoeger2015}. The spectral isolation of the two species ensures that no cooling light impacts the stored qubit \cite{barrett2003}. However, re-cooling is relatively slow and it is desirable to reduce the excitation during transport and separation operations - indeed re-cooling has been the dominant timescale in several prominent demonstrations of the QCCD architecture \cite{home2009, yong2019, pino2021, ryan2021}. 

Although primarily considered here in the context of the QCCD approach and quantum computing, control of mixed species is also important for spectroscopy, where a control ion is manipulated and cooled by laser light and co-trapped with a spectroscopy species of interest \cite{schmidt2005, king2022, schwegler2023}. For spectroscopy species which are challenging to prepare or maintain, transport and separation may facilitate control. An explicit example is the use of transport and separation as part of sympathetic cooling protocols for a single (anti-)proton co-trapped with a single beryllium ion \cite{Cornejo_2021}.

\begin{figure}[t]
    \centering
    \includegraphics[width=1.0\columnwidth]{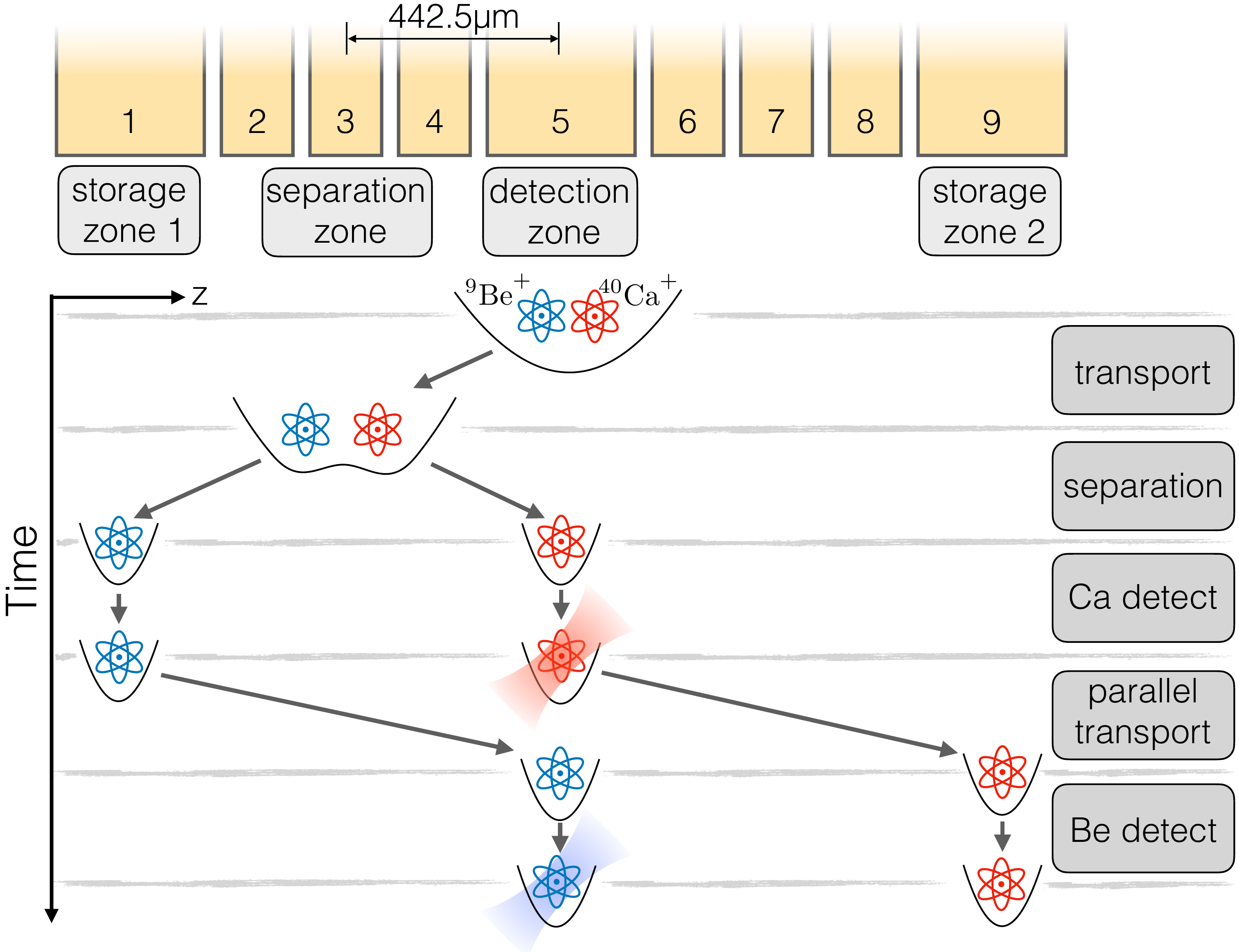}
    \caption{Spatial and temporal schematic of the experiment. One row of segmented DC trap electrodes is shown in yellow. Laser beams for both species are directed through the detection zone. Single \beryllium and \calcium ions are shown as blue and red icons, respectively. The positions of the ions at different steps of the experimental protocol are indicated.
    }
    \label{fig:fig1}
\end{figure}

Transport and separation tasks have been extensively investigated using single species ion chains \cite{rowe2002, leibfried2007, bowler2012, walther2012, ruster2014, kaufmann2014, deClerq2016, sterk2022, tinkey2022, meiners2023, clark2023}, with low excitation achieved over a wide range of parameters. When a second ion species of a different mass is introduced \cite{hanneke2010, negnevitsky2018paper, sosnova2021, burton2023, moses2023race}, the mass-dependent pseudopotential from the radio-frequency trapping fields introduces extra control challenges. The relative participation of each species to the radial normal modes of motion can be highly unbalanced, and different radial modes have very different frequencies \cite{home2013}. In addition, stray radial electric fields displace the ions by different amounts, twisting the ion chain relative to the trap axis and changing the direction along which normal modes are defined. While these effects are present whenever two different mass ions are used, they are exacerbated by the presence of a high mass ratio.

In this Letter, we demonstrate low-excitation transport and separation of a two-ion crystal composed of one \beryllium and one \calcium ion. We identify and investigate several factors inhibiting performance, such as the presence of stray electric fields and the coupling of axial and radial modes. These are mitigated through multi-position stray electric field compensation, the addition of adjustment quadrupole terms in the potential to control mode crossings, and variation of the timescales of dynamic control. A tailored combination of these measures reduces the axial excitation of separated ions to a mean phonon number $\bar{n}~{<}~1.85$ with separation timescales of 2.5 ms.

We use a segmented three-dimensional ion trap \cite{kienzler2015, supplement} with multiple trap electrodes to which dynamic voltages are applied in the transport and separation protocol. Fig.\ref{fig:fig1} shows a sketch of the relevant electrodes as well as the experimental sequence. The protocol starts with a transport sequence T1 moving the two-ion crystal over a distance of \SI{442.5}{\micro\meter} from the detection zone to the separation zone. The latter zone features narrow electrodes for generating strong higher-order potentials. We then run a separation sequence S1, which evolves the trapping potential from a single-well to a double-well configuration. It involves the use of a time-varying axial potential of the form $\Phi(z,t){=}a(t)z^2 + b(t)z^4$, where $z$ is the co-ordinate of the axial direction.
We first ramp up the applied voltages to achieve the highest possible positive axial quartic term, then reduce the quadratic component from a positive value to a negative one until the ions are split and located in a double well \cite{home2006}. The frequency of the lowest-frequency axial mode of the mixed-species Be--Ca crystal reaches its minimum when $a(t){\simeq}0$, i.e. when the potential is purely quartic - we refer to this in what follows as the critical point. At this point the simulated mode frequencies are $f_{Z1}{=}\SI{0.23}{MHz}$, $f_{Z2}{=}\SI{1.05}{MHz}$, $f_{Y2}{=}\SI{2.1}{MHz}$, $f_{X2}{=}\SI{3.6}{MHz}$, $f_{Y1}{=}\SI{12.79}{MHz}$ and $f_{X1}{=}\SI{14.29}{MHz}$, where Z1/Z2, Y1/Y2 and X1/X2 are the axial and the two radial in-phase/out-of-phase modes, respectively \cite{barrett2003}. Afterward, the two harmonic wells containing one ion each are controlled independently and moved apart to a distance of \SI{865}{\micro\meter}, bringing one back to the detection zone while the second is moved to storage zone 1.  After state diagnosis of ion 1, we perform a third parallel transport waveform PT which moves both potential wells such that ion 2 is brought to the detection zone, after which we perform a state diagnosis on this ion. 

Prior to each run of the experiment, we deterministically initialize the ions in the order Be--Ca (\beryllium left, \calcium right) using established re-ordering techniques \cite{home2013}.
The ion crystal is laser-cooled by applying Doppler cooling and EIT cooling \cite{Lechner2016} of all vibrational modes. The EIT cooling detuning is set to optimize cooling of the Y2 mode and also cools the X2 mode \cite{lin2013}, resulting in mean occupancies $\bar{n}_{Y2}{=}\SI{0.14{\pm}0.06}{}$ and $\bar{n}_{X2}{=}\SI{0.36{\pm}0.07}{}$. The X1 and Y1 modes have higher frequencies and are Doppler cooled to $\bar{n}\sim 1$ using beryllium light prior to each sequence. For transport and separation, we focus on the axial modes, since these have participation ratios suitable for multi-qubit gates and sympathetic cooling \cite{negnevitsky2018, pino2021}. Following Doppler and EIT cooling, we therefore sideband cool both the axial Z1 and Z2 modes close to the ground state, resulting in occupations of $\bar{n}_{Z1}{=}\SI{0.03{\pm}0.02}{}$, $\bar{n}_{Z2}{=}\SI{0.01{\pm}0.02}{}$.

Transport and separation waveforms previously developed \cite{marinelli2020} allow us to routinely perform separation of single-species chains of two \beryllium or two \calcium ions with an excitation close to $\Delta\bar{n}_{Be} < 0.8$ and $\Delta\bar{n}_{Ca} < 2.3$ phonons for the single-ion axial vibrational modes respectively (example measurements in the Supplemental Material (SM) \cite{supplement}).
However, the direct use of the same transport and separation waveforms for the mixed-species chain produced large excitation of the axial modes with $\bar{n} \gg \SI{10}{~\phonons}$, which was so severe that unwanted re-ordering of the crystal occurred even during the initial transport segment. A first correction step is to compensate for uncontrolled electric fields radially displacing the ions in a mass-dependent manner \cite{home2013}. The primary excitations are mitigated by adding compensation potentials for stray electric fields at $11$ locations along the trap axis during transport, and at $12$ points in the separation zone throughout the separation waveform. Each of these is calibrated to a level of $\SI{2}{\volt\per\meter}$. 

\begin{figure}[t!]
    \centering
    \includegraphics{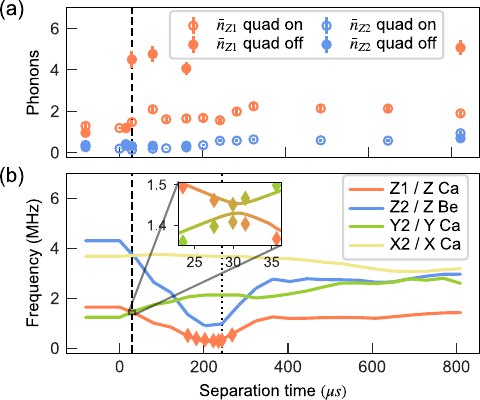}
    \caption{(a) Time-resolved measurement of the axial phonon populations after partially running the separation sequence forward and backwards. $t = 0$ indicates the start of the separation segment, with negative times belonging to the transport waveform. (b) Simulated (solid lines) and measured (diamond points) mode frequencies of the lowest four modes throughout the waveform. Uncertainties on the measured data points are smaller than the marker size. The position of Z1/Y2 mode crossing at \SI{30}{\micro\second} is indicated by the vertical black dashed line. After the vertical dotted line at \SI{243}{\micro\second} the ion separation is greater than \SI{29}{\micro\metre} and we consider the normal modes to be uncoupled. The inset shows measurements and simulations close to the avoided crossing between Z1 and Y2. The color change indicates the transition between the two modes.}
    \label{fig:fig2}
\end{figure}
This compensation allows us to reduce the crystal excitation to a level where no re-ordering occurs during T1, and where diagnostics based on sideband spectroscopy can be used to optimize further. Our approach is to run the sequence up to a point of interest, and then reverse it after a short delay, after which the phonon population is characterized.
Fig.\ref{fig:fig2} shows the measured phonon occupancies of Z1 and Z2 throughout the separation sequence S1, together with the simulated frequencies of Z1, Z2, Y2, and X2 and measured frequencies of Z1 and Y2 for some relevant points. The total duration of the S1 waveform in this example is \SI{823}{\micro\second}, which was the timescale used for our initial diagnostics. The data points at negative times show the contribution of the transport waveform T1, which results in mode occupations of $\bar{n}_{\rm Z1}{\simeq}\num{1.17\pm 0.11}$ and $\bar{n}_{\rm Z2}{\simeq}\num{0.18 \pm 0.09}$ quanta after this part of the protocol. We maintain the Y2 mode at $\SI{230}{\kilo\hertz}$ lower than the Z1 mode throughout T1, and additionally maximize the frequencies of both modes to reduce heating rates.

Due to stray electric fields along the axis, applying a pre-designed waveform for S1 does not necessarily end up with the ions in separate potential wells. Thus we first calibrate the stray field by scanning an additional axial field and observing the transition in populations of the final wells between two ions in the left well, one in each well, and two in the right well. For ion separation we then proceed to use a calibrated value in the middle of the range which results in one ion in each well. 
We observe daily fluctuations of this parameter of about \SI{1}{\volt\per\meter}, which do not affect the performance significantly. Recalibration is only necessary on a timescale of weeks. The temperatures measured after such a calibration are shown in Fig.\ref{fig:fig2} for the separation waveform S1. We observe a significant excitation in the Z1 mode occurring at \SI{30}{\micro\second}, which corresponds in the simulation to a frequency crossing of the Z1 mode with the Y2 mode, highlighted with a black dashed line. Under the conditions of our transport waveform, this is inevitable in separation since the axial curvature is reduced significantly. In our setup, the Y2 mode has a high heating rate of ${\sim}\SI{3000}{\phonons\per\second}$, and thus despite being initially cooled down close to the motional ground state, it gains an excitation of ${\sim}\SI{4}{\phonons}$ by the time the crossing point is reached. The excitation of Z1 then results from the injection of radial phonons due to a coupling between the modes close to the degeneracy point, an effect which has previously been observed in junction transport with a single ion \cite{blakestad2011}. Measurements of the mode frequencies around the crossing point (shown in the inset of Fig.\ref{fig:fig2} (b)), confirm the presence of an avoided crossing with a  minimum mode separation, $\Delta/(2 \pi) {\simeq}\SI{23}{\kilo\hertz}$.
The components of the Hessian relevant to the two modes can be described as ${-}\frac{\hbar \Delta}{2}\hat{\sigma_x}{-}\frac{\hbar \varepsilon(t)}{2}\hat{\sigma_z}$, where $\sigma_x$ and $\sigma_z$ are the Pauli matrices and $\varepsilon(t)$ is the mode separation of the uncoupled modes. We can use the Landau-Zener-Stückelberg–Majorana formula \cite{ivakhnenko2023} to estimate that the probability that the system will undergo a diabatic transition when evolving through the avoided crossing is $P_{D}{=}\exp\left\{{-\Delta^2 /\big[4 \,|\partial / \partial t (f_{Z1} - f_{Y2})|\,\big]}\right\}$. Using the measured $\Delta$ and our characteristic protocol speed we obtain $P_D{=}0.7$. We note that to reduce the exchange of phonons between the modes, the probability of a diabatic transition has to be maximized.

In the single-species case, an off-diagonal coupling term in the Hessian requires the presence of a tilted electric quadrupole, while for a mixed-species chain, such terms can also occur due to a radial electric field, which tilts the ion chain and changes the normal mode expansion along the trap axis. Such a radial field is minimized via our stray field compensation. 
To further suppress the off-diagonal terms, we add to $S1(t)$ a quadrupole potential $\alpha Q(t)$ using an electrode configuration illustrated in Fig.\ref{fig:fig3} (a), where $Q(t)$ is a dimensionless quadrupole term and $\alpha$ is the scaling factor. We empirically scan $\alpha$ to minimize the excitation of the axial mode of \calcium at the end of the separation. Fig.\ref{fig:fig3} shows results of such a calibration along with the measured trap frequencies. We observe that the minimum mode separation close to the avoided crossing point is reduced to ${\simeq}\SI{11}{\kilo\hertz}$, which corresponds to $P_D {=}0.99$.  For the reverse waveform, we recalibrate $\alpha$ by minimizing the excitation of Z1. The mode excitation of Z1 observed after passing forward and backward through the avoided crossing is reduced from \SI{4.29\pm 0.31} quanta to \SI{1.44\pm 0.1} quanta. For mode Z2, we notice no appreciable increase of phonons after the avoided crossings with Y2 and X2 independent of the value of $\alpha$. We attribute this to the fact that the Y2 and X2 modes are strongly dominated by Ca, which in Z2 has a low participation ratio. 

For transport and splitting operations performed on longer ion chains, the mode trajectories exhibit a large number of crossings. Therefore, creating crossing-free waveforms becomes increasingly hard, justifying the need for the techniques developed above. In our particular case, the small size of the ion crystal allows us to generate new transport and separation waveforms T2 and S2 which avoid mode degeneracies by keeping $f_{Y2}{>}f_{Z1} + \SI{380}{\kilo\hertz}$ at all times - these produced similar results to T1-S1 with an optimized quadrupole potential.

\begin{figure}[htb]
    \begin{subfigure}{\columnwidth}
        \centering
        \includegraphics[width=1\linewidth]{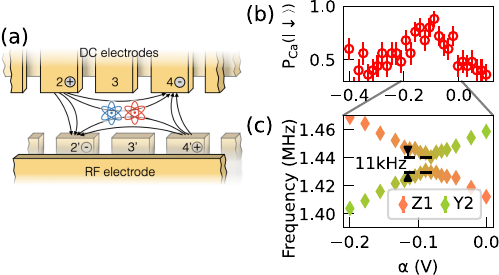}
    \end{subfigure}%
    \caption{Application of a quadrupole potential. (a) Schematic showing the signs of the shifts to the electrode voltages used to generate the quadrupole potential. (b) Calibration of the quadrupole scaling factor $\alpha$ for T1-S1. The measured spin population is a proxy for mode excitation for which a high value indicates low excitation. (c) Measurement of the Z1/Y2 mode frequencies as a function of $\alpha$ close to the avoided mode crossing.}
    \label{fig:fig3}
\end{figure}

\begin{figure}[b]
\centering
\includegraphics[width=\linewidth]{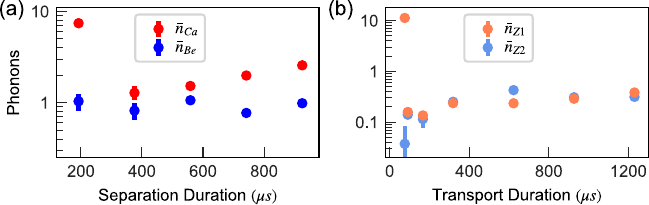}
\caption{\label{fig:fig4}
(a) Excitation of axial modes after T2-S2-PT scanning the total time of S2. (b) Excitation of axial modes after the transport protocol T2-T2$^{-1}$. Excitation is plotted as a function of the total protocol time.}
\end{figure}

The amount of time spent crossing the critical point also strongly affects mode excitation, due to the scaling of the heating rate $\dot{\bar{n}}$ with the trapping frequency.
In a characterization measurement with one \calcium ion, we measure this scaling for the axial mode in the detection zone and find $\dot{\bar{n}} \propto 1/f_{Z}^{5.8}$ -- we suspect this is due to increased technical noise leaking past our filters at low frequencies. Using this value, we estimate the heating rate of mode Z1 at the measured critical point frequency of $\SI{284}{\kilo\hertz}$ to be \SI{2600}{quanta\per\second}. Therefore, we aim to minimize the time spent at the critical point. A contrasting requirement is the desire to satisfy the adiabatic condition $\delta=\left|\frac{1}{2\pi f^2}\frac{d f}{dt}\right|{\ll}1$ throughout transport and separation to avoid coherent excitation of the ion chain \cite{chen2010, bowler2012, ruster2014}. As a result, we optimize the speed at which we run the separation sequence T2-S2-PT - results for the S2 segment are shown in Fig.\ref{fig:fig4}(a). We observe optimal performance at a duration of \SI{375}{\micro\second} ($\delta \sim 0.015$). At shorter timescales ($\delta \sim 0.03$)  we notice that the motional populations show a combination of coherent and thermal fractions, which signifies non-adiabatic excitation \cite{bowler2012}. Scanning the timescales of the transport segments T2-T2$^{-1}$ produced results shown in Fig.\ref{fig:fig4}(b), with a minimum transport time for the \SI{442}{\micro\meter} distance of \SI{46}{\micro\second}. This corresponds to an average velocity of \SI{9.7}{\meter\per\second} without diabatic excitation.

\begin{table}[t]
\begin{tabular}{cccc}
\hline\hline
Sequence                        & Duration      & $\bar{n}_{Z1}$      & $\bar{n}_{Z2}$      \\ \hline
T1 - T1$^{-1}$                  & \SI{1.4}{ms}  & \num{0.82 \pm 0.08} & \num{0.36 \pm 0.07} \\ \hline
T2 - T2$^{-1}$                  & \SI{1.34}{ms} & \num{0.46 \pm 0.03} & \num{0.59 \pm 0.07} \\ \hline
T1 - S1 - S1$^{-1}$ - T1$^{-1}$ & \SI{3.2}{ms}  & \num{2.26 \pm 0.23} & \num{0.98 \pm 0.11} \\ \hline
T2 - S2 - S2$^{-1}$ - T2$^{-1}$ & \SI{2.8}{ms}  &  \num{3.01 \pm 0.2} & \num{1.12 \pm 0.13} \\
\hline\hline
Sequence                        & Duration      & $\bar{n}_{Ca}$      & $\bar{n}_{Be}$      \\ \hline
T1 - S1 - PT                    & \SI{2.7}{ms}  & \num{1.85 \pm 0.15} & \num{0.79 \pm 0.09} \\ \hline
T2 - S2 - PT                    & \SI{2.5}{ms}  & \num{1.40 \pm 0.08} & \num{1.44 \pm 0.09} \\
\hline\hline
\end{tabular}
\caption{
    \label{tab:nbar}
    Mean phonon number in the axial modes of motion for different transport/separation sequences. The upper 4 are for forward and back operation,  probing mode Z1 using \calcium sidebands and mode Z2 using \beryllium. The lower 2 are mean axial excitations of the calcium and beryllium ions, with the latter measured after the PT segment, which has a duration of \SI{1.1}{\milli\second} and results in excitation of \SI{0.15\pm 0.02}{quanta}. The sequence duration is the time for running the full waveform sequence listed.
}
\end{table}

A summary of the performance of all waveforms is given in Tab.\ref{tab:nbar}, including results of separation waveforms run in one direction only as well as forward and back sequences. We observe that the additional quadrupole allows the T1-S1 sequence to achieve similar performance to T2-S2, while the lowest excitations observed are close to 1.5 quanta for \calcium and 1 quantum for \beryllium. The sequence T2-S2-S2$^{-1}$-T2$^{-1}$ shows an excitation that is higher than the one-directional T2-S2-PT for \calcium, with the Z1 mode being heated the most, as would be expected given that this low-frequency mode would be most affected by heating. 

We also performed separation experiments with the reverse ion order, Ca--Be. Measured excitations showed no substantial difference after T2-T2$^{-1}$, but higher final excitations for \calcium were observed after the separation sequence T2-S2-PT. Measurements of the heating rates of modes close to the critical point revealed much higher values for Ca--Be vs. Be--Ca, for reasons which we do not understand (these measurements are presented in the SM \cite{supplement}).

The demonstrated multi-species transport and separation provides a starting point for protocols using more extensive ion arrays. As a first probe of this we apply the waveforms used for two-ion separation to Ca--Be--Ca, Be--Ca--Be, and Ca--Be--Ca--Be crystals \cite{supplement}. While we do not investigate final temperatures for these larger crystals using sideband spectroscopy, we observe that the separation and recombination protocol causes no loss of fluorescence for Ca--Be--Ca, compatible with low motional excitation, and only a small loss for the other configurations. For all crystals, the ion order is maintained. 
The demonstrated low-excitation splitting could be sped up and the heating reduced by engineering non-adiabatic operations \cite{chen2010, bowler2012, ruster2014}. By focusing on the control of trapping fields, our methods are independent of the employed species or trap design involved. They can be readily applied to reduce the required cooling time in other QCCD experiments with different mass ratios \cite{moses2023race, jost_entangled_2009}. In quantum logic spectroscopy protocols, where the mass ratios can be large, the developed techniques are expected to play an even more significant role. As an example, in \cite{Cornejo_2021} the spectroscopy signal degrades with a higher anti-proton temperature, which is affected by imperfect transport and separation operations. Due to the large mass ratio, our techniques could provide a way to keep the temperature low and thereby reduce measurement uncertainties.

FL, SW, and MS performed the mixed-species experiments, while the single-species results were obtained by FL, TB, MM, and VN, with experimental support by BN. Waveform generation was performed by CM, VN, and FL. Data analysis and theoretical understanding of the results were developed by FL, MS, SW, CM, and JH. The paper was written by SW, FL, MS, CM, and JH with input from all authors.

This work was supported by the 
Intelligence Advanced Research Projects Activity (IARPA) via the US Army Research Office grant W911NF-16-1-0070, as well as the  Swiss National Science Foundation (SNF) under Grant No.  $200020\_179147$  and as a part of the NCCR QSIT, a National Centre of Competence (or Excellence) in Research (grant number 51NF40-185902). S.W. acknowledges financial support from the SNSF Swiss Postdoctoral Fellowship (Project no. TMPFP2$\_$210584).




\begin{thebibliography}{46}%
\makeatletter
\providecommand \@ifxundefined [1]{%
 \@ifx{#1\undefined}
}%
\providecommand \@ifnum [1]{%
 \ifnum #1\expandafter \@firstoftwo
 \else \expandafter \@secondoftwo
 \fi
}%
\providecommand \@ifx [1]{%
 \ifx #1\expandafter \@firstoftwo
 \else \expandafter \@secondoftwo
 \fi
}%
\providecommand \natexlab [1]{#1}%
\providecommand \enquote  [1]{``#1''}%
\providecommand \bibnamefont  [1]{#1}%
\providecommand \bibfnamefont [1]{#1}%
\providecommand \citenamefont [1]{#1}%
\providecommand \href@noop [0]{\@secondoftwo}%
\providecommand \href [0]{\begingroup \@sanitize@url \@href}%
\providecommand \@href[1]{\@@startlink{#1}\@@href}%
\providecommand \@@href[1]{\endgroup#1\@@endlink}%
\providecommand \@sanitize@url [0]{\catcode `\\12\catcode `\$12\catcode
  `\&12\catcode `\#12\catcode `\^12\catcode `\_12\catcode `\%12\relax}%
\providecommand \@@startlink[1]{}%
\providecommand \@@endlink[0]{}%
\providecommand \url  [0]{\begingroup\@sanitize@url \@url }%
\providecommand \@url [1]{\endgroup\@href {#1}{\urlprefix }}%
\providecommand \urlprefix  [0]{URL }%
\providecommand \Eprint [0]{\href }%
\providecommand \doibase [0]{http://dx.doi.org/}%
\providecommand \selectlanguage [0]{\@gobble}%
\providecommand \bibinfo  [0]{\@secondoftwo}%
\providecommand \bibfield  [0]{\@secondoftwo}%
\providecommand \translation [1]{[#1]}%
\providecommand \BibitemOpen [0]{}%
\providecommand \bibitemStop [0]{}%
\providecommand \bibitemNoStop [0]{.\EOS\space}%
\providecommand \EOS [0]{\spacefactor3000\relax}%
\providecommand \BibitemShut  [1]{\csname bibitem#1\endcsname}%
\let\auto@bib@innerbib\@empty
\bibitem [{\citenamefont {Clark}\ \emph {et~al.}(2023)\citenamefont {Clark},
  \citenamefont {Herold}, \citenamefont {Merrill}, \citenamefont {Tinkey},
  \citenamefont {Rellergert}, \citenamefont {Clark}, \citenamefont {Brown},
  \citenamefont {Robertson}, \citenamefont {Volin}, \citenamefont {Maller},
  \citenamefont {Shappert}, \citenamefont {McMahon}, \citenamefont {Sawyer},\
  and\ \citenamefont {Brown}}]{clark2023}%
  \BibitemOpen
  \bibfield  {author} {\bibinfo {author} {\bibfnamefont {C.~R.}\ \bibnamefont
  {Clark}}, \bibinfo {author} {\bibfnamefont {C.~D.}\ \bibnamefont {Herold}},
  \bibinfo {author} {\bibfnamefont {J.~T.}\ \bibnamefont {Merrill}}, \bibinfo
  {author} {\bibfnamefont {H.~N.}\ \bibnamefont {Tinkey}}, \bibinfo {author}
  {\bibfnamefont {W.}~\bibnamefont {Rellergert}}, \bibinfo {author}
  {\bibfnamefont {R.}~\bibnamefont {Clark}}, \bibinfo {author} {\bibfnamefont
  {R.}~\bibnamefont {Brown}}, \bibinfo {author} {\bibfnamefont {W.~D.}\
  \bibnamefont {Robertson}}, \bibinfo {author} {\bibfnamefont {C.}~\bibnamefont
  {Volin}}, \bibinfo {author} {\bibfnamefont {K.}~\bibnamefont {Maller}},
  \bibinfo {author} {\bibfnamefont {C.}~\bibnamefont {Shappert}}, \bibinfo
  {author} {\bibfnamefont {B.~J.}\ \bibnamefont {McMahon}}, \bibinfo {author}
  {\bibfnamefont {B.~C.}\ \bibnamefont {Sawyer}}, \ and\ \bibinfo {author}
  {\bibfnamefont {K.~R.}\ \bibnamefont {Brown}},\ }\href
  {https://link.aps.org/doi/10.1103/PhysRevA.107.043119} {\bibfield  {journal}
  {\bibinfo  {journal} {Phys. Rev. A}\ }\textbf {\bibinfo {volume} {107}},\
  \bibinfo {pages} {043119} (\bibinfo {year} {2023})}\BibitemShut {NoStop}%
\bibitem [{\citenamefont {Wang}\ \emph {et~al.}(2021)\citenamefont {Wang},
  \citenamefont {Luan}, \citenamefont {Qiao}, \citenamefont {Um}, \citenamefont
  {Zhang}, \citenamefont {Wang}, \citenamefont {Yuan}, \citenamefont {Gu},
  \citenamefont {Zhang},\ and\ \citenamefont {Kim}}]{wang2021}%
  \BibitemOpen
  \bibfield  {author} {\bibinfo {author} {\bibfnamefont {P.}~\bibnamefont
  {Wang}}, \bibinfo {author} {\bibfnamefont {C.-Y.}\ \bibnamefont {Luan}},
  \bibinfo {author} {\bibfnamefont {M.}~\bibnamefont {Qiao}}, \bibinfo {author}
  {\bibfnamefont {M.}~\bibnamefont {Um}}, \bibinfo {author} {\bibfnamefont
  {J.}~\bibnamefont {Zhang}}, \bibinfo {author} {\bibfnamefont
  {Y.}~\bibnamefont {Wang}}, \bibinfo {author} {\bibfnamefont {X.}~\bibnamefont
  {Yuan}}, \bibinfo {author} {\bibfnamefont {M.}~\bibnamefont {Gu}}, \bibinfo
  {author} {\bibfnamefont {J.}~\bibnamefont {Zhang}}, \ and\ \bibinfo {author}
  {\bibfnamefont {K.}~\bibnamefont {Kim}},\ }\href
  {https://www.nature.com/articles/s41467-020-20330-w} {\bibfield  {journal}
  {\bibinfo  {journal} {Nature communications}\ }\textbf {\bibinfo {volume}
  {12}},\ \bibinfo {pages} {233} (\bibinfo {year} {2021})}\BibitemShut
  {NoStop}%
\bibitem [{\citenamefont {Wineland}\ \emph {et~al.}(1998)\citenamefont
  {Wineland}, \citenamefont {Monroe}, \citenamefont {Itano}, \citenamefont
  {Leibfried}, \citenamefont {King},\ and\ \citenamefont
  {Meekhof}}]{wineland1998}%
  \BibitemOpen
  \bibfield  {author} {\bibinfo {author} {\bibfnamefont {D.~J.}\ \bibnamefont
  {Wineland}}, \bibinfo {author} {\bibfnamefont {C.}~\bibnamefont {Monroe}},
  \bibinfo {author} {\bibfnamefont {W.~M.}\ \bibnamefont {Itano}}, \bibinfo
  {author} {\bibfnamefont {D.}~\bibnamefont {Leibfried}}, \bibinfo {author}
  {\bibfnamefont {B.~E.}\ \bibnamefont {King}}, \ and\ \bibinfo {author}
  {\bibfnamefont {D.~M.}\ \bibnamefont {Meekhof}},\ }\href
  {https://tf.nist.gov/general/pdf/1275.pdf} {\bibfield  {journal} {\bibinfo
  {journal} {Journal of research of the National Institute of Standards and
  Technology}\ }\textbf {\bibinfo {volume} {103}},\ \bibinfo {pages} {259}
  (\bibinfo {year} {1998})}\BibitemShut {NoStop}%
\bibitem [{\citenamefont {Kielpinski}\ \emph {et~al.}(2002)\citenamefont
  {Kielpinski}, \citenamefont {Monroe},\ and\ \citenamefont
  {Wineland}}]{kielpinski2002}%
  \BibitemOpen
  \bibfield  {author} {\bibinfo {author} {\bibfnamefont {D.}~\bibnamefont
  {Kielpinski}}, \bibinfo {author} {\bibfnamefont {C.}~\bibnamefont {Monroe}},
  \ and\ \bibinfo {author} {\bibfnamefont {D.~J.}\ \bibnamefont {Wineland}},\
  }\href {\doibase https://doi.org/10.1038/nature00784} {\bibfield  {journal}
  {\bibinfo  {journal} {Nature}\ }\textbf {\bibinfo {volume} {417}},\ \bibinfo
  {pages} {709} (\bibinfo {year} {2002})}\BibitemShut {NoStop}%
\bibitem [{\citenamefont {Bruzewicz}\ \emph {et~al.}(2019)\citenamefont
  {Bruzewicz}, \citenamefont {Chiaverini}, \citenamefont {McConnell},\ and\
  \citenamefont {Sage}}]{bruzewicz2019}%
  \BibitemOpen
  \bibfield  {author} {\bibinfo {author} {\bibfnamefont {C.~D.}\ \bibnamefont
  {Bruzewicz}}, \bibinfo {author} {\bibfnamefont {J.}~\bibnamefont
  {Chiaverini}}, \bibinfo {author} {\bibfnamefont {R.}~\bibnamefont
  {McConnell}}, \ and\ \bibinfo {author} {\bibfnamefont {J.~M.}\ \bibnamefont
  {Sage}},\ }\href {https://aip.scitation.org/doi/full/10.1063/1.5088164}
  {\bibfield  {journal} {\bibinfo  {journal} {Applied Physics Reviews}\
  }\textbf {\bibinfo {volume} {6}},\ \bibinfo {pages} {021314} (\bibinfo {year}
  {2019})}\BibitemShut {NoStop}%
\bibitem [{\citenamefont {Kaushal}\ \emph {et~al.}(2020)\citenamefont
  {Kaushal}, \citenamefont {Lekitsch}, \citenamefont {Stahl}, \citenamefont
  {Hilder}, \citenamefont {Pijn}, \citenamefont {Schmiegelow}, \citenamefont
  {Bermudez}, \citenamefont {M{\"u}ller}, \citenamefont {Schmidt-Kaler},\ and\
  \citenamefont {Poschinger}}]{kaushal2020}%
  \BibitemOpen
  \bibfield  {author} {\bibinfo {author} {\bibfnamefont {V.}~\bibnamefont
  {Kaushal}}, \bibinfo {author} {\bibfnamefont {B.}~\bibnamefont {Lekitsch}},
  \bibinfo {author} {\bibfnamefont {A.}~\bibnamefont {Stahl}}, \bibinfo
  {author} {\bibfnamefont {J.}~\bibnamefont {Hilder}}, \bibinfo {author}
  {\bibfnamefont {D.}~\bibnamefont {Pijn}}, \bibinfo {author} {\bibfnamefont
  {C.}~\bibnamefont {Schmiegelow}}, \bibinfo {author} {\bibfnamefont
  {A.}~\bibnamefont {Bermudez}}, \bibinfo {author} {\bibfnamefont
  {M.}~\bibnamefont {M{\"u}ller}}, \bibinfo {author} {\bibfnamefont
  {F.}~\bibnamefont {Schmidt-Kaler}}, \ and\ \bibinfo {author} {\bibfnamefont
  {U.}~\bibnamefont {Poschinger}},\ }\href
  {https://avs.scitation.org/doi/full/10.1116/1.5126186} {\bibfield  {journal}
  {\bibinfo  {journal} {AVS Quantum Science}\ }\textbf {\bibinfo {volume}
  {2}},\ \bibinfo {pages} {014101} (\bibinfo {year} {2020})}\BibitemShut
  {NoStop}%
\bibitem [{\citenamefont {H{\"a}ffner}\ \emph {et~al.}(2008)\citenamefont
  {H{\"a}ffner}, \citenamefont {Roos},\ and\ \citenamefont
  {Blatt}}]{haffner2008}%
  \BibitemOpen
  \bibfield  {author} {\bibinfo {author} {\bibfnamefont {H.}~\bibnamefont
  {H{\"a}ffner}}, \bibinfo {author} {\bibfnamefont {C.~F.}\ \bibnamefont
  {Roos}}, \ and\ \bibinfo {author} {\bibfnamefont {R.}~\bibnamefont {Blatt}},\
  }\href {https://www.sciencedirect.com/science/article/pii/S0370157308003463}
  {\bibfield  {journal} {\bibinfo  {journal} {Physics reports}\ }\textbf
  {\bibinfo {volume} {469}},\ \bibinfo {pages} {155} (\bibinfo {year}
  {2008})}\BibitemShut {NoStop}%
\bibitem [{\citenamefont {Schm{\"o}ger}\ \emph {et~al.}(2015)\citenamefont
  {Schm{\"o}ger}, \citenamefont {Versolato}, \citenamefont {Schwarz},
  \citenamefont {Kohnen}, \citenamefont {Windberger}, \citenamefont {Piest},
  \citenamefont {Feuchtenbeiner}, \citenamefont {Pedregosa-Gutierrez},
  \citenamefont {Leopold}, \citenamefont {Micke}, \citenamefont {Hansen},
  \citenamefont {Baumann}, \citenamefont {Drewsen}, \citenamefont {Ullrich}, ,
  \citenamefont {Schmidt},\ and\ \citenamefont {Crespo
  L\'{o}pez-Urrutia}}]{schmoeger2015}%
  \BibitemOpen
  \bibfield  {author} {\bibinfo {author} {\bibfnamefont {L.}~\bibnamefont
  {Schm{\"o}ger}}, \bibinfo {author} {\bibfnamefont {O.}~\bibnamefont
  {Versolato}}, \bibinfo {author} {\bibfnamefont {M.}~\bibnamefont {Schwarz}},
  \bibinfo {author} {\bibfnamefont {M.}~\bibnamefont {Kohnen}}, \bibinfo
  {author} {\bibfnamefont {A.}~\bibnamefont {Windberger}}, \bibinfo {author}
  {\bibfnamefont {B.}~\bibnamefont {Piest}}, \bibinfo {author} {\bibfnamefont
  {S.}~\bibnamefont {Feuchtenbeiner}}, \bibinfo {author} {\bibfnamefont
  {J.}~\bibnamefont {Pedregosa-Gutierrez}}, \bibinfo {author} {\bibfnamefont
  {T.}~\bibnamefont {Leopold}}, \bibinfo {author} {\bibfnamefont
  {P.}~\bibnamefont {Micke}}, \bibinfo {author} {\bibfnamefont
  {A.}~\bibnamefont {Hansen}}, \bibinfo {author} {\bibfnamefont
  {T.}~\bibnamefont {Baumann}}, \bibinfo {author} {\bibfnamefont
  {M.}~\bibnamefont {Drewsen}}, \bibinfo {author} {\bibfnamefont
  {J.}~\bibnamefont {Ullrich}}, , \bibinfo {author} {\bibfnamefont {P.~O.}\
  \bibnamefont {Schmidt}}, \ and\ \bibinfo {author} {\bibfnamefont {J.~R.}\
  \bibnamefont {Crespo L\'{o}pez-Urrutia}},\ }\href
  {https://www.science.org/doi/10.1126/science.aaa2960} {\bibfield  {journal}
  {\bibinfo  {journal} {Science}\ }\textbf {\bibinfo {volume} {347}},\ \bibinfo
  {pages} {1233} (\bibinfo {year} {2015})}\BibitemShut {NoStop}%
\bibitem [{\citenamefont {Barrett}\ \emph {et~al.}(2003)\citenamefont
  {Barrett}, \citenamefont {DeMarco}, \citenamefont {Schaetz}, \citenamefont
  {Meyer}, \citenamefont {Leibfried}, \citenamefont {Britton}, \citenamefont
  {Chiaverini}, \citenamefont {Itano}, \citenamefont
  {Jelenkovi\ifmmode~\acute{c}\else \'{c}\fi{}}, \citenamefont {Jost},
  \citenamefont {Langer}, \citenamefont {Rosenband},\ and\ \citenamefont
  {Wineland}}]{barrett2003}%
  \BibitemOpen
  \bibfield  {author} {\bibinfo {author} {\bibfnamefont {M.~D.}\ \bibnamefont
  {Barrett}}, \bibinfo {author} {\bibfnamefont {B.}~\bibnamefont {DeMarco}},
  \bibinfo {author} {\bibfnamefont {T.}~\bibnamefont {Schaetz}}, \bibinfo
  {author} {\bibfnamefont {V.}~\bibnamefont {Meyer}}, \bibinfo {author}
  {\bibfnamefont {D.}~\bibnamefont {Leibfried}}, \bibinfo {author}
  {\bibfnamefont {J.}~\bibnamefont {Britton}}, \bibinfo {author} {\bibfnamefont
  {J.}~\bibnamefont {Chiaverini}}, \bibinfo {author} {\bibfnamefont {W.~M.}\
  \bibnamefont {Itano}}, \bibinfo {author} {\bibfnamefont {B.}~\bibnamefont
  {Jelenkovi\ifmmode~\acute{c}\else \'{c}\fi{}}}, \bibinfo {author}
  {\bibfnamefont {J.~D.}\ \bibnamefont {Jost}}, \bibinfo {author}
  {\bibfnamefont {C.}~\bibnamefont {Langer}}, \bibinfo {author} {\bibfnamefont
  {T.}~\bibnamefont {Rosenband}}, \ and\ \bibinfo {author} {\bibfnamefont
  {D.~J.}\ \bibnamefont {Wineland}},\ }\href {\doibase
  10.1103/PhysRevA.68.042302} {\bibfield  {journal} {\bibinfo  {journal} {Phys.
  Rev. A}\ }\textbf {\bibinfo {volume} {68}},\ \bibinfo {pages} {042302}
  (\bibinfo {year} {2003})}\BibitemShut {NoStop}%
\bibitem [{\citenamefont {Home}\ \emph {et~al.}(2009)\citenamefont {Home},
  \citenamefont {Hanneke}, \citenamefont {Jost}, \citenamefont {Amini},
  \citenamefont {Leibfried},\ and\ \citenamefont {Wineland}}]{home2009}%
  \BibitemOpen
  \bibfield  {author} {\bibinfo {author} {\bibfnamefont {J.~P.}\ \bibnamefont
  {Home}}, \bibinfo {author} {\bibfnamefont {D.}~\bibnamefont {Hanneke}},
  \bibinfo {author} {\bibfnamefont {J.~D.}\ \bibnamefont {Jost}}, \bibinfo
  {author} {\bibfnamefont {J.~M.}\ \bibnamefont {Amini}}, \bibinfo {author}
  {\bibfnamefont {D.}~\bibnamefont {Leibfried}}, \ and\ \bibinfo {author}
  {\bibfnamefont {D.~J.}\ \bibnamefont {Wineland}},\ }\href
  {https://www.science.org/doi/full/10.1126/science.1177077} {\bibfield
  {journal} {\bibinfo  {journal} {Science}\ }\textbf {\bibinfo {volume}
  {325}},\ \bibinfo {pages} {1227} (\bibinfo {year} {2009})}\BibitemShut
  {NoStop}%
\bibitem [{\citenamefont {Wan}\ \emph {et~al.}(2019)\citenamefont {Wan},
  \citenamefont {Kienzler}, \citenamefont {Erickson}, \citenamefont {Mayer},
  \citenamefont {Tan}, \citenamefont {Wu}, \citenamefont {Vasconcelos},
  \citenamefont {Glancy}, \citenamefont {Knill}, \citenamefont {Wineland},
  \citenamefont {Wilson},\ and\ \citenamefont {Leibfried}}]{yong2019}%
  \BibitemOpen
  \bibfield  {author} {\bibinfo {author} {\bibfnamefont {Y.}~\bibnamefont
  {Wan}}, \bibinfo {author} {\bibfnamefont {D.}~\bibnamefont {Kienzler}},
  \bibinfo {author} {\bibfnamefont {S.~D.}\ \bibnamefont {Erickson}}, \bibinfo
  {author} {\bibfnamefont {K.~H.}\ \bibnamefont {Mayer}}, \bibinfo {author}
  {\bibfnamefont {T.~R.}\ \bibnamefont {Tan}}, \bibinfo {author} {\bibfnamefont
  {J.~J.}\ \bibnamefont {Wu}}, \bibinfo {author} {\bibfnamefont {H.~M.}\
  \bibnamefont {Vasconcelos}}, \bibinfo {author} {\bibfnamefont
  {S.}~\bibnamefont {Glancy}}, \bibinfo {author} {\bibfnamefont
  {E.}~\bibnamefont {Knill}}, \bibinfo {author} {\bibfnamefont {D.~J.}\
  \bibnamefont {Wineland}}, \bibinfo {author} {\bibfnamefont {A.~C.}\
  \bibnamefont {Wilson}}, \ and\ \bibinfo {author} {\bibfnamefont
  {D.}~\bibnamefont {Leibfried}},\ }\href {\doibase 10.1126/science.aaw9415}
  {\bibfield  {journal} {\bibinfo  {journal} {Science}\ }\textbf {\bibinfo
  {volume} {364}},\ \bibinfo {pages} {875} (\bibinfo {year}
  {2019})}\BibitemShut {NoStop}%
\bibitem [{\citenamefont {Pino}\ \emph {et~al.}(2021)\citenamefont {Pino},
  \citenamefont {Dreiling}, \citenamefont {Figgatt}, \citenamefont {Gaebler},
  \citenamefont {Moses}, \citenamefont {Allman}, \citenamefont {Baldwin},
  \citenamefont {Foss-Feig}, \citenamefont {Hayes}, \citenamefont {Mayer} \emph
  {et~al.}}]{pino2021}%
  \BibitemOpen
  \bibfield  {author} {\bibinfo {author} {\bibfnamefont {J.~M.}\ \bibnamefont
  {Pino}}, \bibinfo {author} {\bibfnamefont {J.~M.}\ \bibnamefont {Dreiling}},
  \bibinfo {author} {\bibfnamefont {C.}~\bibnamefont {Figgatt}}, \bibinfo
  {author} {\bibfnamefont {J.~P.}\ \bibnamefont {Gaebler}}, \bibinfo {author}
  {\bibfnamefont {S.~A.}\ \bibnamefont {Moses}}, \bibinfo {author}
  {\bibfnamefont {M.}~\bibnamefont {Allman}}, \bibinfo {author} {\bibfnamefont
  {C.}~\bibnamefont {Baldwin}}, \bibinfo {author} {\bibfnamefont
  {M.}~\bibnamefont {Foss-Feig}}, \bibinfo {author} {\bibfnamefont
  {D.}~\bibnamefont {Hayes}}, \bibinfo {author} {\bibfnamefont
  {K.}~\bibnamefont {Mayer}},  \emph {et~al.},\ }\href
  {https://www.nature.com/articles/s41586-021-03318-4} {\bibfield  {journal}
  {\bibinfo  {journal} {Nature}\ }\textbf {\bibinfo {volume} {592}},\ \bibinfo
  {pages} {209} (\bibinfo {year} {2021})}\BibitemShut {NoStop}%
\bibitem [{\citenamefont {Ryan-Anderson}\ \emph {et~al.}(2021)\citenamefont
  {Ryan-Anderson}, \citenamefont {Bohnet}, \citenamefont {Lee}, \citenamefont
  {Gresh}, \citenamefont {Hankin}, \citenamefont {Gaebler}, \citenamefont
  {Francois}, \citenamefont {Chernoguzov}, \citenamefont {Lucchetti},
  \citenamefont {Brown}, \citenamefont {Gatterman}, \citenamefont {Halit},
  \citenamefont {Gilmore}, \citenamefont {Gerber}, \citenamefont {Neyenhuis},
  \citenamefont {Hayes},\ and\ \citenamefont {Stutz}}]{ryan2021}%
  \BibitemOpen
  \bibfield  {author} {\bibinfo {author} {\bibfnamefont {C.}~\bibnamefont
  {Ryan-Anderson}}, \bibinfo {author} {\bibfnamefont {J.~G.}\ \bibnamefont
  {Bohnet}}, \bibinfo {author} {\bibfnamefont {K.}~\bibnamefont {Lee}},
  \bibinfo {author} {\bibfnamefont {D.}~\bibnamefont {Gresh}}, \bibinfo
  {author} {\bibfnamefont {A.}~\bibnamefont {Hankin}}, \bibinfo {author}
  {\bibfnamefont {J.~P.}\ \bibnamefont {Gaebler}}, \bibinfo {author}
  {\bibfnamefont {D.}~\bibnamefont {Francois}}, \bibinfo {author}
  {\bibfnamefont {A.}~\bibnamefont {Chernoguzov}}, \bibinfo {author}
  {\bibfnamefont {D.}~\bibnamefont {Lucchetti}}, \bibinfo {author}
  {\bibfnamefont {N.~C.}\ \bibnamefont {Brown}}, \bibinfo {author}
  {\bibfnamefont {T.~M.}\ \bibnamefont {Gatterman}}, \bibinfo {author}
  {\bibfnamefont {S.~K.}\ \bibnamefont {Halit}}, \bibinfo {author}
  {\bibfnamefont {K.}~\bibnamefont {Gilmore}}, \bibinfo {author} {\bibfnamefont
  {J.~A.}\ \bibnamefont {Gerber}}, \bibinfo {author} {\bibfnamefont
  {B.}~\bibnamefont {Neyenhuis}}, \bibinfo {author} {\bibfnamefont
  {D.}~\bibnamefont {Hayes}}, \ and\ \bibinfo {author} {\bibfnamefont {R.~P.}\
  \bibnamefont {Stutz}},\ }\href {\doibase 10.1103/PhysRevX.11.041058}
  {\bibfield  {journal} {\bibinfo  {journal} {Phys. Rev. X}\ }\textbf {\bibinfo
  {volume} {11}},\ \bibinfo {pages} {041058} (\bibinfo {year}
  {2021})}\BibitemShut {NoStop}%
\bibitem [{\citenamefont {Schmidt}\ \emph {et~al.}(2005)\citenamefont
  {Schmidt}, \citenamefont {Rosenband}, \citenamefont {Langer}, \citenamefont
  {Itano}, \citenamefont {Bergquist},\ and\ \citenamefont
  {Wineland}}]{schmidt2005}%
  \BibitemOpen
  \bibfield  {author} {\bibinfo {author} {\bibfnamefont {P.~O.}\ \bibnamefont
  {Schmidt}}, \bibinfo {author} {\bibfnamefont {T.}~\bibnamefont {Rosenband}},
  \bibinfo {author} {\bibfnamefont {C.}~\bibnamefont {Langer}}, \bibinfo
  {author} {\bibfnamefont {W.~M.}\ \bibnamefont {Itano}}, \bibinfo {author}
  {\bibfnamefont {J.~C.}\ \bibnamefont {Bergquist}}, \ and\ \bibinfo {author}
  {\bibfnamefont {D.~J.}\ \bibnamefont {Wineland}},\ }\href
  {https://www.science.org/doi/10.1126/science.1114375} {\bibfield  {journal}
  {\bibinfo  {journal} {Science}\ }\textbf {\bibinfo {volume} {309}},\ \bibinfo
  {pages} {749} (\bibinfo {year} {2005})}\BibitemShut {NoStop}%
\bibitem [{\citenamefont {King}\ \emph {et~al.}(2022)\citenamefont {King},
  \citenamefont {Spie{\ss}}, \citenamefont {Micke}, \citenamefont {Wilzewski},
  \citenamefont {Leopold}, \citenamefont {Benkler}, \citenamefont {Lange},
  \citenamefont {Huntemann}, \citenamefont {Surzhykov}, \citenamefont
  {Yerokhin}, \citenamefont {L\'{o}pez-Urrutia},\ and\ \citenamefont
  {Schmidt}}]{king2022}%
  \BibitemOpen
  \bibfield  {author} {\bibinfo {author} {\bibfnamefont {S.~A.}\ \bibnamefont
  {King}}, \bibinfo {author} {\bibfnamefont {L.~J.}\ \bibnamefont {Spie{\ss}}},
  \bibinfo {author} {\bibfnamefont {P.}~\bibnamefont {Micke}}, \bibinfo
  {author} {\bibfnamefont {A.}~\bibnamefont {Wilzewski}}, \bibinfo {author}
  {\bibfnamefont {T.}~\bibnamefont {Leopold}}, \bibinfo {author} {\bibfnamefont
  {E.}~\bibnamefont {Benkler}}, \bibinfo {author} {\bibfnamefont
  {R.}~\bibnamefont {Lange}}, \bibinfo {author} {\bibfnamefont
  {N.}~\bibnamefont {Huntemann}}, \bibinfo {author} {\bibfnamefont
  {A.}~\bibnamefont {Surzhykov}}, \bibinfo {author} {\bibfnamefont {V.~A.}\
  \bibnamefont {Yerokhin}}, \bibinfo {author} {\bibfnamefont {J.~R.~C.}\
  \bibnamefont {L\'{o}pez-Urrutia}}, \ and\ \bibinfo {author} {\bibfnamefont
  {P.~O.}\ \bibnamefont {Schmidt}},\ }\href
  {https://www.nature.com/articles/s41586-022-05245-4} {\bibfield  {journal}
  {\bibinfo  {journal} {Nature}\ }\textbf {\bibinfo {volume} {611}},\ \bibinfo
  {pages} {43} (\bibinfo {year} {2022})}\BibitemShut {NoStop}%
\bibitem [{\citenamefont {Schwegler}\ \emph {et~al.}(2023)\citenamefont
  {Schwegler}, \citenamefont {Holzapfel}, \citenamefont {Stadler},
  \citenamefont {Mitjans}, \citenamefont {Sergachev}, \citenamefont {Home},\
  and\ \citenamefont {Kienzler}}]{schwegler2023}%
  \BibitemOpen
  \bibfield  {author} {\bibinfo {author} {\bibfnamefont {N.}~\bibnamefont
  {Schwegler}}, \bibinfo {author} {\bibfnamefont {D.}~\bibnamefont
  {Holzapfel}}, \bibinfo {author} {\bibfnamefont {M.}~\bibnamefont {Stadler}},
  \bibinfo {author} {\bibfnamefont {A.}~\bibnamefont {Mitjans}}, \bibinfo
  {author} {\bibfnamefont {I.}~\bibnamefont {Sergachev}}, \bibinfo {author}
  {\bibfnamefont {J.~P.}\ \bibnamefont {Home}}, \ and\ \bibinfo {author}
  {\bibfnamefont {D.}~\bibnamefont {Kienzler}},\ }\href
  {https://link.aps.org/doi/10.1103/PhysRevLett.131.133003} {\bibfield
  {journal} {\bibinfo  {journal} {Phys. Rev. Lett.}\ }\textbf {\bibinfo
  {volume} {131}},\ \bibinfo {pages} {133003} (\bibinfo {year}
  {2023})}\BibitemShut {NoStop}%
\bibitem [{\citenamefont {Cornejo}\ \emph {et~al.}(2021)\citenamefont
  {Cornejo}, \citenamefont {Lehnert}, \citenamefont {Niemann}, \citenamefont
  {Mielke}, \citenamefont {Meiners}, \citenamefont {Bautista-Salvador},
  \citenamefont {Schulte}, \citenamefont {Nitzschke}, \citenamefont {Borchert},
  \citenamefont {Hammerer}, \citenamefont {Ulmer},\ and\ \citenamefont
  {Ospelkaus}}]{Cornejo_2021}%
  \BibitemOpen
  \bibfield  {author} {\bibinfo {author} {\bibfnamefont {J.~M.}\ \bibnamefont
  {Cornejo}}, \bibinfo {author} {\bibfnamefont {R.}~\bibnamefont {Lehnert}},
  \bibinfo {author} {\bibfnamefont {M.}~\bibnamefont {Niemann}}, \bibinfo
  {author} {\bibfnamefont {J.}~\bibnamefont {Mielke}}, \bibinfo {author}
  {\bibfnamefont {T.}~\bibnamefont {Meiners}}, \bibinfo {author} {\bibfnamefont
  {A.}~\bibnamefont {Bautista-Salvador}}, \bibinfo {author} {\bibfnamefont
  {M.}~\bibnamefont {Schulte}}, \bibinfo {author} {\bibfnamefont
  {D.}~\bibnamefont {Nitzschke}}, \bibinfo {author} {\bibfnamefont {M.~J.}\
  \bibnamefont {Borchert}}, \bibinfo {author} {\bibfnamefont {K.}~\bibnamefont
  {Hammerer}}, \bibinfo {author} {\bibfnamefont {S.}~\bibnamefont {Ulmer}}, \
  and\ \bibinfo {author} {\bibfnamefont {C.}~\bibnamefont {Ospelkaus}},\ }\href
  {\doibase 10.1088/1367-2630/ac136e} {\bibfield  {journal} {\bibinfo
  {journal} {New Journal of Physics}\ }\textbf {\bibinfo {volume} {23}},\
  \bibinfo {pages} {073045} (\bibinfo {year} {2021})}\BibitemShut {NoStop}%
\bibitem [{\citenamefont {Rowe}\ \emph {et~al.}(2002)\citenamefont {Rowe},
  \citenamefont {Ben-Kish}, \citenamefont {Demarco}, \citenamefont {Leibfried},
  \citenamefont {Meyer}, \citenamefont {Beall}, \citenamefont {Britton},
  \citenamefont {Hughes}, \citenamefont {Itano}, \citenamefont {Jelenkovic}
  \emph {et~al.}}]{rowe2002}%
  \BibitemOpen
  \bibfield  {author} {\bibinfo {author} {\bibfnamefont {M.~A.}\ \bibnamefont
  {Rowe}}, \bibinfo {author} {\bibfnamefont {A.}~\bibnamefont {Ben-Kish}},
  \bibinfo {author} {\bibfnamefont {B.}~\bibnamefont {Demarco}}, \bibinfo
  {author} {\bibfnamefont {D.}~\bibnamefont {Leibfried}}, \bibinfo {author}
  {\bibfnamefont {V.}~\bibnamefont {Meyer}}, \bibinfo {author} {\bibfnamefont
  {J.}~\bibnamefont {Beall}}, \bibinfo {author} {\bibfnamefont
  {J.}~\bibnamefont {Britton}}, \bibinfo {author} {\bibfnamefont
  {J.}~\bibnamefont {Hughes}}, \bibinfo {author} {\bibfnamefont {W.~M.}\
  \bibnamefont {Itano}}, \bibinfo {author} {\bibfnamefont {B.}~\bibnamefont
  {Jelenkovic}},  \emph {et~al.},\ }\href
  {https://arxiv.org/abs/quant-ph/0205094} {\bibfield  {journal} {\bibinfo
  {journal} {arXiv preprint quant-ph/0205094}\ } (\bibinfo {year}
  {2002})}\BibitemShut {NoStop}%
\bibitem [{\citenamefont {Leibfried}\ \emph {et~al.}(2007)\citenamefont
  {Leibfried}, \citenamefont {Knill}, \citenamefont {Ospelkaus},\ and\
  \citenamefont {Wineland}}]{leibfried2007}%
  \BibitemOpen
  \bibfield  {author} {\bibinfo {author} {\bibfnamefont {D.}~\bibnamefont
  {Leibfried}}, \bibinfo {author} {\bibfnamefont {E.}~\bibnamefont {Knill}},
  \bibinfo {author} {\bibfnamefont {C.}~\bibnamefont {Ospelkaus}}, \ and\
  \bibinfo {author} {\bibfnamefont {D.~J.}\ \bibnamefont {Wineland}},\ }\href
  {https://journals.aps.org/pra/abstract/10.1103/PhysRevA.76.032324} {\bibfield
   {journal} {\bibinfo  {journal} {Physical Review A}\ }\textbf {\bibinfo
  {volume} {76}},\ \bibinfo {pages} {032324} (\bibinfo {year}
  {2007})}\BibitemShut {NoStop}%
\bibitem [{\citenamefont {Bowler}\ \emph {et~al.}(2012)\citenamefont {Bowler},
  \citenamefont {Gaebler}, \citenamefont {Lin}, \citenamefont {Tan},
  \citenamefont {Hanneke}, \citenamefont {Jost}, \citenamefont {Home},
  \citenamefont {Leibfried},\ and\ \citenamefont {Wineland}}]{bowler2012}%
  \BibitemOpen
  \bibfield  {author} {\bibinfo {author} {\bibfnamefont {R.}~\bibnamefont
  {Bowler}}, \bibinfo {author} {\bibfnamefont {J.}~\bibnamefont {Gaebler}},
  \bibinfo {author} {\bibfnamefont {Y.}~\bibnamefont {Lin}}, \bibinfo {author}
  {\bibfnamefont {T.~R.}\ \bibnamefont {Tan}}, \bibinfo {author} {\bibfnamefont
  {D.}~\bibnamefont {Hanneke}}, \bibinfo {author} {\bibfnamefont {J.~D.}\
  \bibnamefont {Jost}}, \bibinfo {author} {\bibfnamefont {J.}~\bibnamefont
  {Home}}, \bibinfo {author} {\bibfnamefont {D.}~\bibnamefont {Leibfried}}, \
  and\ \bibinfo {author} {\bibfnamefont {D.~J.}\ \bibnamefont {Wineland}},\
  }\href {https://journals.aps.org/prl/abstract/10.1103/PhysRevLett.109.080502}
  {\bibfield  {journal} {\bibinfo  {journal} {Physical Review Letters}\
  }\textbf {\bibinfo {volume} {109}},\ \bibinfo {pages} {080502} (\bibinfo
  {year} {2012})}\BibitemShut {NoStop}%
\bibitem [{\citenamefont {Walther}\ \emph {et~al.}(2012)\citenamefont
  {Walther}, \citenamefont {Ziesel}, \citenamefont {Ruster}, \citenamefont
  {Dawkins}, \citenamefont {Ott}, \citenamefont {Hettrich}, \citenamefont
  {Singer}, \citenamefont {Schmidt-Kaler},\ and\ \citenamefont
  {Poschinger}}]{walther2012}%
  \BibitemOpen
  \bibfield  {author} {\bibinfo {author} {\bibfnamefont {A.}~\bibnamefont
  {Walther}}, \bibinfo {author} {\bibfnamefont {F.}~\bibnamefont {Ziesel}},
  \bibinfo {author} {\bibfnamefont {T.}~\bibnamefont {Ruster}}, \bibinfo
  {author} {\bibfnamefont {S.~T.}\ \bibnamefont {Dawkins}}, \bibinfo {author}
  {\bibfnamefont {K.}~\bibnamefont {Ott}}, \bibinfo {author} {\bibfnamefont
  {M.}~\bibnamefont {Hettrich}}, \bibinfo {author} {\bibfnamefont
  {K.}~\bibnamefont {Singer}}, \bibinfo {author} {\bibfnamefont
  {F.}~\bibnamefont {Schmidt-Kaler}}, \ and\ \bibinfo {author} {\bibfnamefont
  {U.}~\bibnamefont {Poschinger}},\ }\href
  {https://journals.aps.org/prl/abstract/10.1103/PhysRevLett.109.080501}
  {\bibfield  {journal} {\bibinfo  {journal} {Physical Review Letters}\
  }\textbf {\bibinfo {volume} {109}},\ \bibinfo {pages} {080501} (\bibinfo
  {year} {2012})}\BibitemShut {NoStop}%
\bibitem [{\citenamefont {Ruster}\ \emph {et~al.}(2014)\citenamefont {Ruster},
  \citenamefont {Warschburger}, \citenamefont {Kaufmann}, \citenamefont
  {Schmiegelow}, \citenamefont {Walther}, \citenamefont {Hettrich},
  \citenamefont {Pfister}, \citenamefont {Kaushal}, \citenamefont
  {Schmidt-Kaler},\ and\ \citenamefont {Poschinger}}]{ruster2014}%
  \BibitemOpen
  \bibfield  {author} {\bibinfo {author} {\bibfnamefont {T.}~\bibnamefont
  {Ruster}}, \bibinfo {author} {\bibfnamefont {C.}~\bibnamefont
  {Warschburger}}, \bibinfo {author} {\bibfnamefont {H.}~\bibnamefont
  {Kaufmann}}, \bibinfo {author} {\bibfnamefont {C.~T.}\ \bibnamefont
  {Schmiegelow}}, \bibinfo {author} {\bibfnamefont {A.}~\bibnamefont
  {Walther}}, \bibinfo {author} {\bibfnamefont {M.}~\bibnamefont {Hettrich}},
  \bibinfo {author} {\bibfnamefont {A.}~\bibnamefont {Pfister}}, \bibinfo
  {author} {\bibfnamefont {V.}~\bibnamefont {Kaushal}}, \bibinfo {author}
  {\bibfnamefont {F.}~\bibnamefont {Schmidt-Kaler}}, \ and\ \bibinfo {author}
  {\bibfnamefont {U.~G.}\ \bibnamefont {Poschinger}},\ }\href
  {https://journals.aps.org/pra/abstract/10.1103/PhysRevA.90.033410} {\bibfield
   {journal} {\bibinfo  {journal} {Physical Review A}\ }\textbf {\bibinfo
  {volume} {90}},\ \bibinfo {pages} {033410} (\bibinfo {year}
  {2014})}\BibitemShut {NoStop}%
\bibitem [{\citenamefont {Kaufmann}\ \emph {et~al.}(2014)\citenamefont
  {Kaufmann}, \citenamefont {Ruster}, \citenamefont {Schmiegelow},
  \citenamefont {Schmidt-Kaler},\ and\ \citenamefont
  {Poschinger}}]{kaufmann2014}%
  \BibitemOpen
  \bibfield  {author} {\bibinfo {author} {\bibfnamefont {H.}~\bibnamefont
  {Kaufmann}}, \bibinfo {author} {\bibfnamefont {T.}~\bibnamefont {Ruster}},
  \bibinfo {author} {\bibfnamefont {C.}~\bibnamefont {Schmiegelow}}, \bibinfo
  {author} {\bibfnamefont {F.}~\bibnamefont {Schmidt-Kaler}}, \ and\ \bibinfo
  {author} {\bibfnamefont {U.}~\bibnamefont {Poschinger}},\ }\href
  {https://iopscience.iop.org/article/10.1088/1367-2630/16/7/073012} {\bibfield
   {journal} {\bibinfo  {journal} {New Journal of Physics}\ }\textbf {\bibinfo
  {volume} {16}},\ \bibinfo {pages} {073012} (\bibinfo {year}
  {2014})}\BibitemShut {NoStop}%
\bibitem [{\citenamefont {de~Clercq}\ \emph {et~al.}(2016)\citenamefont
  {de~Clercq}, \citenamefont {Lo}, \citenamefont {Marinelli}, \citenamefont
  {Nadlinger}, \citenamefont {Oswald}, \citenamefont {Negnevitsky},
  \citenamefont {Kienzler}, \citenamefont {Keitch},\ and\ \citenamefont
  {Home}}]{deClerq2016}%
  \BibitemOpen
  \bibfield  {author} {\bibinfo {author} {\bibfnamefont {L.~E.}\ \bibnamefont
  {de~Clercq}}, \bibinfo {author} {\bibfnamefont {H.-Y.}\ \bibnamefont {Lo}},
  \bibinfo {author} {\bibfnamefont {M.}~\bibnamefont {Marinelli}}, \bibinfo
  {author} {\bibfnamefont {D.}~\bibnamefont {Nadlinger}}, \bibinfo {author}
  {\bibfnamefont {R.}~\bibnamefont {Oswald}}, \bibinfo {author} {\bibfnamefont
  {V.}~\bibnamefont {Negnevitsky}}, \bibinfo {author} {\bibfnamefont
  {D.}~\bibnamefont {Kienzler}}, \bibinfo {author} {\bibfnamefont
  {B.}~\bibnamefont {Keitch}}, \ and\ \bibinfo {author} {\bibfnamefont {J.~P.}\
  \bibnamefont {Home}},\ }\href
  {https://journals.aps.org/prl/abstract/10.1103/PhysRevLett.116.080502}
  {\bibfield  {journal} {\bibinfo  {journal} {Physical Review Letters}\
  }\textbf {\bibinfo {volume} {116}},\ \bibinfo {pages} {080502} (\bibinfo
  {year} {2016})}\BibitemShut {NoStop}%
\bibitem [{\citenamefont {Sterk}\ \emph {et~al.}(2022)\citenamefont {Sterk},
  \citenamefont {Coakley}, \citenamefont {Goldberg}, \citenamefont {Hietala},
  \citenamefont {Lechtenberg}, \citenamefont {McGuinness}, \citenamefont
  {McMurtrey}, \citenamefont {Parazzoli}, \citenamefont {Van Der~Wall},\ and\
  \citenamefont {Stick}}]{sterk2022}%
  \BibitemOpen
  \bibfield  {author} {\bibinfo {author} {\bibfnamefont {J.~D.}\ \bibnamefont
  {Sterk}}, \bibinfo {author} {\bibfnamefont {H.}~\bibnamefont {Coakley}},
  \bibinfo {author} {\bibfnamefont {J.}~\bibnamefont {Goldberg}}, \bibinfo
  {author} {\bibfnamefont {V.}~\bibnamefont {Hietala}}, \bibinfo {author}
  {\bibfnamefont {J.}~\bibnamefont {Lechtenberg}}, \bibinfo {author}
  {\bibfnamefont {H.}~\bibnamefont {McGuinness}}, \bibinfo {author}
  {\bibfnamefont {D.}~\bibnamefont {McMurtrey}}, \bibinfo {author}
  {\bibfnamefont {L.~P.}\ \bibnamefont {Parazzoli}}, \bibinfo {author}
  {\bibfnamefont {J.}~\bibnamefont {Van Der~Wall}}, \ and\ \bibinfo {author}
  {\bibfnamefont {D.}~\bibnamefont {Stick}},\ }\href
  {https://www.nature.com/articles/s41534-022-00579-3} {\bibfield  {journal}
  {\bibinfo  {journal} {npj Quantum Information}\ }\textbf {\bibinfo {volume}
  {8}},\ \bibinfo {pages} {68} (\bibinfo {year} {2022})}\BibitemShut {NoStop}%
\bibitem [{\citenamefont {Tinkey}\ \emph {et~al.}(2022)\citenamefont {Tinkey},
  \citenamefont {Clark}, \citenamefont {Sawyer},\ and\ \citenamefont
  {Brown}}]{tinkey2022}%
  \BibitemOpen
  \bibfield  {author} {\bibinfo {author} {\bibfnamefont {H.~N.}\ \bibnamefont
  {Tinkey}}, \bibinfo {author} {\bibfnamefont {C.~R.}\ \bibnamefont {Clark}},
  \bibinfo {author} {\bibfnamefont {B.~C.}\ \bibnamefont {Sawyer}}, \ and\
  \bibinfo {author} {\bibfnamefont {K.~R.}\ \bibnamefont {Brown}},\ }\href
  {https://journals.aps.org/prl/abstract/10.1103/PhysRevLett.128.050502}
  {\bibfield  {journal} {\bibinfo  {journal} {Physical Review Letters}\
  }\textbf {\bibinfo {volume} {128}},\ \bibinfo {pages} {050502} (\bibinfo
  {year} {2022})}\BibitemShut {NoStop}%
\bibitem [{\citenamefont {Meiners}\ \emph {et~al.}(2023)\citenamefont
  {Meiners}, \citenamefont {Coenders}, \citenamefont {Mielke}, \citenamefont
  {Niemann}, \citenamefont {Cornejo}, \citenamefont {Ulmer},\ and\
  \citenamefont {Ospelkaus}}]{meiners2023}%
  \BibitemOpen
  \bibfield  {author} {\bibinfo {author} {\bibfnamefont {T.}~\bibnamefont
  {Meiners}}, \bibinfo {author} {\bibfnamefont {J.-A.}\ \bibnamefont
  {Coenders}}, \bibinfo {author} {\bibfnamefont {J.}~\bibnamefont {Mielke}},
  \bibinfo {author} {\bibfnamefont {M.}~\bibnamefont {Niemann}}, \bibinfo
  {author} {\bibfnamefont {J.}~\bibnamefont {Cornejo}}, \bibinfo {author}
  {\bibfnamefont {S.}~\bibnamefont {Ulmer}}, \ and\ \bibinfo {author}
  {\bibfnamefont {C.}~\bibnamefont {Ospelkaus}},\ }\href
  {https://arxiv.org/abs/2309.06776} {\bibfield  {journal} {\bibinfo  {journal}
  {arXiv preprint arXiv:2309.06776}\ } (\bibinfo {year} {2023})}\BibitemShut
  {NoStop}%
\bibitem [{\citenamefont {Hanneke}\ \emph {et~al.}(2010)\citenamefont
  {Hanneke}, \citenamefont {Home}, \citenamefont {Jost}, \citenamefont {Amini},
  \citenamefont {Leibfried},\ and\ \citenamefont {Wineland}}]{hanneke2010}%
  \BibitemOpen
  \bibfield  {author} {\bibinfo {author} {\bibfnamefont {D.}~\bibnamefont
  {Hanneke}}, \bibinfo {author} {\bibfnamefont {J.}~\bibnamefont {Home}},
  \bibinfo {author} {\bibfnamefont {J.~D.}\ \bibnamefont {Jost}}, \bibinfo
  {author} {\bibfnamefont {J.~M.}\ \bibnamefont {Amini}}, \bibinfo {author}
  {\bibfnamefont {D.}~\bibnamefont {Leibfried}}, \ and\ \bibinfo {author}
  {\bibfnamefont {D.~J.}\ \bibnamefont {Wineland}},\ }\href
  {https://www.nature.com/articles/nphys1453} {\bibfield  {journal} {\bibinfo
  {journal} {Nature Physics}\ }\textbf {\bibinfo {volume} {6}},\ \bibinfo
  {pages} {13} (\bibinfo {year} {2010})}\BibitemShut {NoStop}%
\bibitem [{\citenamefont {Negnevitsky}\ \emph {et~al.}(2018)\citenamefont
  {Negnevitsky}, \citenamefont {Marinelli}, \citenamefont {Mehta},
  \citenamefont {Lo}, \citenamefont {Fl{\"u}hmann},\ and\ \citenamefont
  {Home}}]{negnevitsky2018paper}%
  \BibitemOpen
  \bibfield  {author} {\bibinfo {author} {\bibfnamefont {V.}~\bibnamefont
  {Negnevitsky}}, \bibinfo {author} {\bibfnamefont {M.}~\bibnamefont
  {Marinelli}}, \bibinfo {author} {\bibfnamefont {K.~K.}\ \bibnamefont
  {Mehta}}, \bibinfo {author} {\bibfnamefont {H.-Y.}\ \bibnamefont {Lo}},
  \bibinfo {author} {\bibfnamefont {C.}~\bibnamefont {Fl{\"u}hmann}}, \ and\
  \bibinfo {author} {\bibfnamefont {J.~P.}\ \bibnamefont {Home}},\ }\href
  {\doibase 10.1038/s41586-018-0668-z} {\bibfield  {journal} {\bibinfo
  {journal} {Nature}\ }\textbf {\bibinfo {volume} {563}},\ \bibinfo {pages}
  {527} (\bibinfo {year} {2018})}\BibitemShut {NoStop}%
\bibitem [{\citenamefont {Sosnova}\ \emph {et~al.}(2021)\citenamefont
  {Sosnova}, \citenamefont {Carter},\ and\ \citenamefont
  {Monroe}}]{sosnova2021}%
  \BibitemOpen
  \bibfield  {author} {\bibinfo {author} {\bibfnamefont {K.}~\bibnamefont
  {Sosnova}}, \bibinfo {author} {\bibfnamefont {A.}~\bibnamefont {Carter}}, \
  and\ \bibinfo {author} {\bibfnamefont {C.}~\bibnamefont {Monroe}},\ }\href
  {\doibase 10.1103/PhysRevA.103.012610} {\bibfield  {journal} {\bibinfo
  {journal} {Phys. Rev. A}\ }\textbf {\bibinfo {volume} {103}},\ \bibinfo
  {pages} {012610} (\bibinfo {year} {2021})}\BibitemShut {NoStop}%
\bibitem [{\citenamefont {Burton}\ \emph {et~al.}(2023)\citenamefont {Burton},
  \citenamefont {Estey}, \citenamefont {Hoffman}, \citenamefont {Perry},
  \citenamefont {Volin},\ and\ \citenamefont {Price}}]{burton2023}%
  \BibitemOpen
  \bibfield  {author} {\bibinfo {author} {\bibfnamefont {W.~C.}\ \bibnamefont
  {Burton}}, \bibinfo {author} {\bibfnamefont {B.}~\bibnamefont {Estey}},
  \bibinfo {author} {\bibfnamefont {I.~M.}\ \bibnamefont {Hoffman}}, \bibinfo
  {author} {\bibfnamefont {A.~R.}\ \bibnamefont {Perry}}, \bibinfo {author}
  {\bibfnamefont {C.}~\bibnamefont {Volin}}, \ and\ \bibinfo {author}
  {\bibfnamefont {G.}~\bibnamefont {Price}},\ }\href
  {https://link.aps.org/doi/10.1103/PhysRevLett.130.173202} {\bibfield
  {journal} {\bibinfo  {journal} {Phys. Rev. Lett.}\ }\textbf {\bibinfo
  {volume} {130}},\ \bibinfo {pages} {173202} (\bibinfo {year}
  {2023})}\BibitemShut {NoStop}%
\bibitem [{\citenamefont {Moses}\ \emph {et~al.}(2023)\citenamefont {Moses},
  \citenamefont {Baldwin}, \citenamefont {Allman}, \citenamefont {Ancona},
  \citenamefont {Ascarrunz}, \citenamefont {Barnes}, \citenamefont
  {Bartolotta}, \citenamefont {Bjork}, \citenamefont {Blanchard}, \citenamefont
  {Bohn}, \citenamefont {Bohnet}, \citenamefont {Brown}, \citenamefont
  {Burdick}, \citenamefont {Burton}, \citenamefont {Campbell}, \citenamefont
  {Campora}, \citenamefont {Carron}, \citenamefont {Chambers}, \citenamefont
  {Chan}, \citenamefont {Chen}, \citenamefont {Chernoguzov}, \citenamefont
  {Chertkov}, \citenamefont {Colina}, \citenamefont {Curtis}, \citenamefont
  {Daniel}, \citenamefont {DeCross}, \citenamefont {Deen}, \citenamefont
  {Delaney}, \citenamefont {Dreiling}, \citenamefont {Ertsgaard}, \citenamefont
  {Esposito}, \citenamefont {Estey}, \citenamefont {Fabrikant}, \citenamefont
  {Figgatt}, \citenamefont {Foltz}, \citenamefont {Foss-Feig}, \citenamefont
  {Francois}, \citenamefont {Gaebler}, \citenamefont {Gatterman}, \citenamefont
  {Gilbreth}, \citenamefont {Giles}, \citenamefont {Glynn}, \citenamefont
  {Hall}, \citenamefont {Hankin}, \citenamefont {Hansen}, \citenamefont
  {Hayes}, \citenamefont {Higashi}, \citenamefont {Hoffman}, \citenamefont
  {Horning}, \citenamefont {Hout}, \citenamefont {Jacobs}, \citenamefont
  {Johansen}, \citenamefont {Jones}, \citenamefont {Karcz}, \citenamefont
  {Klein}, \citenamefont {Lauria}, \citenamefont {Lee}, \citenamefont {Liefer},
  \citenamefont {Lu}, \citenamefont {Lucchetti}, \citenamefont {Lytle},
  \citenamefont {Malm}, \citenamefont {Matheny}, \citenamefont {Mathewson},
  \citenamefont {Mayer}, \citenamefont {Miller}, \citenamefont {Mills},
  \citenamefont {Neyenhuis}, \citenamefont {Nugent}, \citenamefont {Olson},
  \citenamefont {Parks}, \citenamefont {Price}, \citenamefont {Price},
  \citenamefont {Pugh}, \citenamefont {Ransford}, \citenamefont {Reed},
  \citenamefont {Roman}, \citenamefont {Rowe}, \citenamefont {Ryan-Anderson},
  \citenamefont {Sanders}, \citenamefont {Sedlacek}, \citenamefont {Shevchuk},
  \citenamefont {Siegfried}, \citenamefont {Skripka}, \citenamefont {Spaun},
  \citenamefont {Sprenkle}, \citenamefont {Stutz}, \citenamefont {Swallows},
  \citenamefont {Tobey}, \citenamefont {Tran}, \citenamefont {Tran},
  \citenamefont {Vogt}, \citenamefont {Volin}, \citenamefont {Walker},
  \citenamefont {Zolot},\ and\ \citenamefont {Pino}}]{moses2023race}%
  \BibitemOpen
  \bibfield  {author} {\bibinfo {author} {\bibfnamefont {S.~A.}\ \bibnamefont
  {Moses}}, \bibinfo {author} {\bibfnamefont {C.~H.}\ \bibnamefont {Baldwin}},
  \bibinfo {author} {\bibfnamefont {M.~S.}\ \bibnamefont {Allman}}, \bibinfo
  {author} {\bibfnamefont {R.}~\bibnamefont {Ancona}}, \bibinfo {author}
  {\bibfnamefont {L.}~\bibnamefont {Ascarrunz}}, \bibinfo {author}
  {\bibfnamefont {C.}~\bibnamefont {Barnes}}, \bibinfo {author} {\bibfnamefont
  {J.}~\bibnamefont {Bartolotta}}, \bibinfo {author} {\bibfnamefont
  {B.}~\bibnamefont {Bjork}}, \bibinfo {author} {\bibfnamefont
  {P.}~\bibnamefont {Blanchard}}, \bibinfo {author} {\bibfnamefont
  {M.}~\bibnamefont {Bohn}}, \bibinfo {author} {\bibfnamefont {J.~G.}\
  \bibnamefont {Bohnet}}, \bibinfo {author} {\bibfnamefont {N.~C.}\
  \bibnamefont {Brown}}, \bibinfo {author} {\bibfnamefont {N.~Q.}\ \bibnamefont
  {Burdick}}, \bibinfo {author} {\bibfnamefont {W.~C.}\ \bibnamefont {Burton}},
  \bibinfo {author} {\bibfnamefont {S.~L.}\ \bibnamefont {Campbell}}, \bibinfo
  {author} {\bibfnamefont {J.~P.}\ \bibnamefont {Campora}}, \bibinfo {author}
  {\bibfnamefont {C.}~\bibnamefont {Carron}}, \bibinfo {author} {\bibfnamefont
  {J.}~\bibnamefont {Chambers}}, \bibinfo {author} {\bibfnamefont {J.~W.}\
  \bibnamefont {Chan}}, \bibinfo {author} {\bibfnamefont {Y.~H.}\ \bibnamefont
  {Chen}}, \bibinfo {author} {\bibfnamefont {A.}~\bibnamefont {Chernoguzov}},
  \bibinfo {author} {\bibfnamefont {E.}~\bibnamefont {Chertkov}}, \bibinfo
  {author} {\bibfnamefont {J.}~\bibnamefont {Colina}}, \bibinfo {author}
  {\bibfnamefont {J.~P.}\ \bibnamefont {Curtis}}, \bibinfo {author}
  {\bibfnamefont {R.}~\bibnamefont {Daniel}}, \bibinfo {author} {\bibfnamefont
  {M.}~\bibnamefont {DeCross}}, \bibinfo {author} {\bibfnamefont
  {D.}~\bibnamefont {Deen}}, \bibinfo {author} {\bibfnamefont {C.}~\bibnamefont
  {Delaney}}, \bibinfo {author} {\bibfnamefont {J.~M.}\ \bibnamefont
  {Dreiling}}, \bibinfo {author} {\bibfnamefont {C.~T.}\ \bibnamefont
  {Ertsgaard}}, \bibinfo {author} {\bibfnamefont {J.}~\bibnamefont {Esposito}},
  \bibinfo {author} {\bibfnamefont {B.}~\bibnamefont {Estey}}, \bibinfo
  {author} {\bibfnamefont {M.}~\bibnamefont {Fabrikant}}, \bibinfo {author}
  {\bibfnamefont {C.}~\bibnamefont {Figgatt}}, \bibinfo {author} {\bibfnamefont
  {C.}~\bibnamefont {Foltz}}, \bibinfo {author} {\bibfnamefont
  {M.}~\bibnamefont {Foss-Feig}}, \bibinfo {author} {\bibfnamefont
  {D.}~\bibnamefont {Francois}}, \bibinfo {author} {\bibfnamefont {J.~P.}\
  \bibnamefont {Gaebler}}, \bibinfo {author} {\bibfnamefont {T.~M.}\
  \bibnamefont {Gatterman}}, \bibinfo {author} {\bibfnamefont {C.~N.}\
  \bibnamefont {Gilbreth}}, \bibinfo {author} {\bibfnamefont {J.}~\bibnamefont
  {Giles}}, \bibinfo {author} {\bibfnamefont {E.}~\bibnamefont {Glynn}},
  \bibinfo {author} {\bibfnamefont {A.}~\bibnamefont {Hall}}, \bibinfo {author}
  {\bibfnamefont {A.~M.}\ \bibnamefont {Hankin}}, \bibinfo {author}
  {\bibfnamefont {A.}~\bibnamefont {Hansen}}, \bibinfo {author} {\bibfnamefont
  {D.}~\bibnamefont {Hayes}}, \bibinfo {author} {\bibfnamefont
  {B.}~\bibnamefont {Higashi}}, \bibinfo {author} {\bibfnamefont {I.~M.}\
  \bibnamefont {Hoffman}}, \bibinfo {author} {\bibfnamefont {B.}~\bibnamefont
  {Horning}}, \bibinfo {author} {\bibfnamefont {J.~J.}\ \bibnamefont {Hout}},
  \bibinfo {author} {\bibfnamefont {R.}~\bibnamefont {Jacobs}}, \bibinfo
  {author} {\bibfnamefont {J.}~\bibnamefont {Johansen}}, \bibinfo {author}
  {\bibfnamefont {L.}~\bibnamefont {Jones}}, \bibinfo {author} {\bibfnamefont
  {J.}~\bibnamefont {Karcz}}, \bibinfo {author} {\bibfnamefont
  {T.}~\bibnamefont {Klein}}, \bibinfo {author} {\bibfnamefont
  {P.}~\bibnamefont {Lauria}}, \bibinfo {author} {\bibfnamefont
  {P.}~\bibnamefont {Lee}}, \bibinfo {author} {\bibfnamefont {D.}~\bibnamefont
  {Liefer}}, \bibinfo {author} {\bibfnamefont {S.~T.}\ \bibnamefont {Lu}},
  \bibinfo {author} {\bibfnamefont {D.}~\bibnamefont {Lucchetti}}, \bibinfo
  {author} {\bibfnamefont {C.}~\bibnamefont {Lytle}}, \bibinfo {author}
  {\bibfnamefont {A.}~\bibnamefont {Malm}}, \bibinfo {author} {\bibfnamefont
  {M.}~\bibnamefont {Matheny}}, \bibinfo {author} {\bibfnamefont
  {B.}~\bibnamefont {Mathewson}}, \bibinfo {author} {\bibfnamefont
  {K.}~\bibnamefont {Mayer}}, \bibinfo {author} {\bibfnamefont {D.~B.}\
  \bibnamefont {Miller}}, \bibinfo {author} {\bibfnamefont {M.}~\bibnamefont
  {Mills}}, \bibinfo {author} {\bibfnamefont {B.}~\bibnamefont {Neyenhuis}},
  \bibinfo {author} {\bibfnamefont {L.}~\bibnamefont {Nugent}}, \bibinfo
  {author} {\bibfnamefont {S.}~\bibnamefont {Olson}}, \bibinfo {author}
  {\bibfnamefont {J.}~\bibnamefont {Parks}}, \bibinfo {author} {\bibfnamefont
  {G.~N.}\ \bibnamefont {Price}}, \bibinfo {author} {\bibfnamefont
  {Z.}~\bibnamefont {Price}}, \bibinfo {author} {\bibfnamefont
  {M.}~\bibnamefont {Pugh}}, \bibinfo {author} {\bibfnamefont {A.}~\bibnamefont
  {Ransford}}, \bibinfo {author} {\bibfnamefont {A.~P.}\ \bibnamefont {Reed}},
  \bibinfo {author} {\bibfnamefont {C.}~\bibnamefont {Roman}}, \bibinfo
  {author} {\bibfnamefont {M.}~\bibnamefont {Rowe}}, \bibinfo {author}
  {\bibfnamefont {C.}~\bibnamefont {Ryan-Anderson}}, \bibinfo {author}
  {\bibfnamefont {S.}~\bibnamefont {Sanders}}, \bibinfo {author} {\bibfnamefont
  {J.}~\bibnamefont {Sedlacek}}, \bibinfo {author} {\bibfnamefont
  {P.}~\bibnamefont {Shevchuk}}, \bibinfo {author} {\bibfnamefont
  {P.}~\bibnamefont {Siegfried}}, \bibinfo {author} {\bibfnamefont
  {T.}~\bibnamefont {Skripka}}, \bibinfo {author} {\bibfnamefont
  {B.}~\bibnamefont {Spaun}}, \bibinfo {author} {\bibfnamefont {R.~T.}\
  \bibnamefont {Sprenkle}}, \bibinfo {author} {\bibfnamefont {R.~P.}\
  \bibnamefont {Stutz}}, \bibinfo {author} {\bibfnamefont {M.}~\bibnamefont
  {Swallows}}, \bibinfo {author} {\bibfnamefont {R.~I.}\ \bibnamefont {Tobey}},
  \bibinfo {author} {\bibfnamefont {A.}~\bibnamefont {Tran}}, \bibinfo {author}
  {\bibfnamefont {T.}~\bibnamefont {Tran}}, \bibinfo {author} {\bibfnamefont
  {E.}~\bibnamefont {Vogt}}, \bibinfo {author} {\bibfnamefont {C.}~\bibnamefont
  {Volin}}, \bibinfo {author} {\bibfnamefont {J.}~\bibnamefont {Walker}},
  \bibinfo {author} {\bibfnamefont {A.~M.}\ \bibnamefont {Zolot}}, \ and\
  \bibinfo {author} {\bibfnamefont {J.~M.}\ \bibnamefont {Pino}},\ }\href
  {\doibase 10.1103/PhysRevX.13.041052} {\bibfield  {journal} {\bibinfo
  {journal} {Phys. Rev. X}\ }\textbf {\bibinfo {volume} {13}},\ \bibinfo
  {pages} {041052} (\bibinfo {year} {2023})}\BibitemShut {NoStop}%
\bibitem [{\citenamefont {Home}(2013)}]{home2013}%
  \BibitemOpen
  \bibfield  {author} {\bibinfo {author} {\bibfnamefont {J.~P.}\ \bibnamefont
  {Home}},\ }in\ \href {\doibase 10.1016/B978-0-12-408090-4.00004-9} {\emph
  {\bibinfo {booktitle} {Advances {{In Atomic}}, {{Molecular}}, and {{Optical
  Physics}}}}},\ \bibinfo {series} {Advances in {{Atomic}}, {{Molecular}}, and
  {{Optical Physics}}}, Vol.~\bibinfo {volume} {62},\ \bibinfo {editor} {edited
  by\ \bibinfo {editor} {\bibfnamefont {E.}~\bibnamefont {Arimondo}}, \bibinfo
  {editor} {\bibfnamefont {P.~R.}\ \bibnamefont {Berman}}, \ and\ \bibinfo
  {editor} {\bibfnamefont {C.~C.}\ \bibnamefont {Lin}}}\ (\bibinfo  {publisher}
  {{Academic Press}},\ \bibinfo {year} {2013})\ pp.\ \bibinfo {pages}
  {231--277}\BibitemShut {NoStop}%
\bibitem [{\citenamefont {Kienzler}(2015)}]{kienzler2015}%
  \BibitemOpen
  \bibfield  {author} {\bibinfo {author} {\bibfnamefont {D.}~\bibnamefont
  {Kienzler}},\ }\emph {\bibinfo {title} {Quantum harmonic oscillator state
  synthesis by reservoir engineering}},\ \href
  {https://www.research-collection.ethz.ch/handle/20.500.11850/102459} {Ph.D.
  thesis},\ \bibinfo  {school} {ETH Zurich} (\bibinfo {year}
  {2015})\BibitemShut {NoStop}%
\bibitem [{sup()}]{supplement}%
  \BibitemOpen
  \href@noop {} {\bibinfo  {journal} {See Supplemental Material for a detailed
  description of the fitting model for the extraction of the phonon population,
  the micromotion compensation procedure, the temporal shape of the applied
  trap voltages, results on the separation of single-species crystals, a
  comparison of separation excitation for Be-Ca and Ca-Be, and the separation
  of larger ion crystals. The Supplemental Material contains Refs.
  [22,24,33,34,39,45,46]}\ }\BibitemShut {NoStop}%
\bibitem [{\citenamefont {Home}\ and\ \citenamefont {Steane}(2006)}]{home2006}%
  \BibitemOpen
\bibfield  {journal} {  }\bibfield  {author} {\bibinfo {author} {\bibfnamefont
  {J.~P.}\ \bibnamefont {Home}}\ and\ \bibinfo {author} {\bibfnamefont {A.~M.}\
  \bibnamefont {Steane}},\ }\href {\doibase 10.48550/arxiv.quant-ph/0411102}
  {\bibfield  {journal} {\bibinfo  {journal} {Quantum Information and
  Computation}\ }\textbf {\bibinfo {volume} {6}} (\bibinfo {year} {2006}),\
  10.48550/arxiv.quant-ph/0411102}\BibitemShut {NoStop}%
\bibitem [{\citenamefont {Lechner}\ \emph {et~al.}(2016)\citenamefont
  {Lechner}, \citenamefont {Maier}, \citenamefont {Hempel}, \citenamefont
  {Jurcevic}, \citenamefont {Lanyon}, \citenamefont {Monz}, \citenamefont
  {Brownnutt}, \citenamefont {Blatt},\ and\ \citenamefont
  {Roos}}]{Lechner2016}%
  \BibitemOpen
  \bibfield  {author} {\bibinfo {author} {\bibfnamefont {R.}~\bibnamefont
  {Lechner}}, \bibinfo {author} {\bibfnamefont {C.}~\bibnamefont {Maier}},
  \bibinfo {author} {\bibfnamefont {C.}~\bibnamefont {Hempel}}, \bibinfo
  {author} {\bibfnamefont {P.}~\bibnamefont {Jurcevic}}, \bibinfo {author}
  {\bibfnamefont {B.~P.}\ \bibnamefont {Lanyon}}, \bibinfo {author}
  {\bibfnamefont {T.}~\bibnamefont {Monz}}, \bibinfo {author} {\bibfnamefont
  {M.}~\bibnamefont {Brownnutt}}, \bibinfo {author} {\bibfnamefont
  {R.}~\bibnamefont {Blatt}}, \ and\ \bibinfo {author} {\bibfnamefont {C.~F.}\
  \bibnamefont {Roos}},\ }\href {\doibase 10.1103/PhysRevA.93.053401}
  {\bibfield  {journal} {\bibinfo  {journal} {Phys. Rev. A}\ }\textbf {\bibinfo
  {volume} {93}},\ \bibinfo {pages} {053401} (\bibinfo {year}
  {2016})}\BibitemShut {NoStop}%
\bibitem [{\citenamefont {Lin}\ \emph {et~al.}(2013)\citenamefont {Lin},
  \citenamefont {Gaebler}, \citenamefont {Tan}, \citenamefont {Bowler},
  \citenamefont {Jost}, \citenamefont {Leibfried},\ and\ \citenamefont
  {Wineland}}]{lin2013}%
  \BibitemOpen
  \bibfield  {author} {\bibinfo {author} {\bibfnamefont {Y.}~\bibnamefont
  {Lin}}, \bibinfo {author} {\bibfnamefont {J.~P.}\ \bibnamefont {Gaebler}},
  \bibinfo {author} {\bibfnamefont {T.~R.}\ \bibnamefont {Tan}}, \bibinfo
  {author} {\bibfnamefont {R.}~\bibnamefont {Bowler}}, \bibinfo {author}
  {\bibfnamefont {J.~D.}\ \bibnamefont {Jost}}, \bibinfo {author}
  {\bibfnamefont {D.}~\bibnamefont {Leibfried}}, \ and\ \bibinfo {author}
  {\bibfnamefont {D.~J.}\ \bibnamefont {Wineland}},\ }\href {\doibase
  10.1103/PhysRevLett.110.153002} {\bibfield  {journal} {\bibinfo  {journal}
  {Phys. Rev. Lett.}\ }\textbf {\bibinfo {volume} {110}},\ \bibinfo {pages}
  {153002} (\bibinfo {year} {2013})}\BibitemShut {NoStop}%
\bibitem [{\citenamefont {Negnevitsky}(2018)}]{negnevitsky2018}%
  \BibitemOpen
  \bibfield  {author} {\bibinfo {author} {\bibfnamefont {V.}~\bibnamefont
  {Negnevitsky}},\ }\emph {\bibinfo {title} {Feedback-stabilised quantum states
  in a mixed-species ion system}},\ \href
  {https://www.research-collection.ethz.ch/handle/20.500.11850/295923} {Ph.D.
  thesis},\ \bibinfo  {school} {ETH Zurich} (\bibinfo {year}
  {2018})\BibitemShut {NoStop}%
\bibitem [{\citenamefont {Marinelli}(2020)}]{marinelli2020}%
  \BibitemOpen
  \bibfield  {author} {\bibinfo {author} {\bibfnamefont {M.}~\bibnamefont
  {Marinelli}},\ }\emph {\bibinfo {title} {Quantum information processing with
  mixed-species ion crystals}},\ \href
  {https://www.research-collection.ethz.ch/handle/20.500.11850/420229} {Ph.D.
  thesis},\ \bibinfo  {school} {ETH Zurich} (\bibinfo {year}
  {2020})\BibitemShut {NoStop}%
\bibitem [{\citenamefont {Blakestad}\ \emph {et~al.}(2011)\citenamefont
  {Blakestad}, \citenamefont {Ospelkaus}, \citenamefont {VanDevender},
  \citenamefont {Wesenberg}, \citenamefont {Biercuk}, \citenamefont
  {Leibfried},\ and\ \citenamefont {Wineland}}]{blakestad2011}%
  \BibitemOpen
  \bibfield  {author} {\bibinfo {author} {\bibfnamefont {R.~B.}\ \bibnamefont
  {Blakestad}}, \bibinfo {author} {\bibfnamefont {C.}~\bibnamefont
  {Ospelkaus}}, \bibinfo {author} {\bibfnamefont {A.~P.}\ \bibnamefont
  {VanDevender}}, \bibinfo {author} {\bibfnamefont {J.~H.}\ \bibnamefont
  {Wesenberg}}, \bibinfo {author} {\bibfnamefont {M.~J.}\ \bibnamefont
  {Biercuk}}, \bibinfo {author} {\bibfnamefont {D.}~\bibnamefont {Leibfried}},
  \ and\ \bibinfo {author} {\bibfnamefont {D.~J.}\ \bibnamefont {Wineland}},\
  }\href {\doibase 10.1103/PhysRevA.84.032314} {\bibfield  {journal} {\bibinfo
  {journal} {Phys. Rev. A}\ }\textbf {\bibinfo {volume} {84}},\ \bibinfo
  {pages} {032314} (\bibinfo {year} {2011})}\BibitemShut {NoStop}%
\bibitem [{\citenamefont {Ivakhnenko}\ \emph {et~al.}(2023)\citenamefont
  {Ivakhnenko}, \citenamefont {Shevchenko},\ and\ \citenamefont
  {Nori}}]{ivakhnenko2023}%
  \BibitemOpen
  \bibfield  {author} {\bibinfo {author} {\bibfnamefont {O.~V.}\ \bibnamefont
  {Ivakhnenko}}, \bibinfo {author} {\bibfnamefont {S.~N.}\ \bibnamefont
  {Shevchenko}}, \ and\ \bibinfo {author} {\bibfnamefont {F.}~\bibnamefont
  {Nori}},\ }\href {\doibase https://doi.org/10.1016/j.physrep.2022.10.002}
  {\bibfield  {journal} {\bibinfo  {journal} {Physics Reports}\ }\textbf
  {\bibinfo {volume} {995}},\ \bibinfo {pages} {1} (\bibinfo {year}
  {2023})}\BibitemShut {NoStop}%
\bibitem [{\citenamefont {Chen}\ \emph {et~al.}(2010)\citenamefont {Chen},
  \citenamefont {Ruschhaupt}, \citenamefont {Schmidt}, \citenamefont {del
  Campo}, \citenamefont {Gu\'ery-Odelin},\ and\ \citenamefont
  {Muga}}]{chen2010}%
  \BibitemOpen
  \bibfield  {author} {\bibinfo {author} {\bibfnamefont {X.}~\bibnamefont
  {Chen}}, \bibinfo {author} {\bibfnamefont {A.}~\bibnamefont {Ruschhaupt}},
  \bibinfo {author} {\bibfnamefont {S.}~\bibnamefont {Schmidt}}, \bibinfo
  {author} {\bibfnamefont {A.}~\bibnamefont {del Campo}}, \bibinfo {author}
  {\bibfnamefont {D.}~\bibnamefont {Gu\'ery-Odelin}}, \ and\ \bibinfo {author}
  {\bibfnamefont {J.~G.}\ \bibnamefont {Muga}},\ }\href
  {https://link.aps.org/doi/10.1103/PhysRevLett.104.063002} {\bibfield
  {journal} {\bibinfo  {journal} {Phys. Rev. Lett.}\ }\textbf {\bibinfo
  {volume} {104}},\ \bibinfo {pages} {063002} (\bibinfo {year}
  {2010})}\BibitemShut {NoStop}%
\bibitem [{\citenamefont {Jost}\ \emph {et~al.}(2009)\citenamefont {Jost},
  \citenamefont {Home}, \citenamefont {Amini}, \citenamefont {Hanneke},
  \citenamefont {Ozeri}, \citenamefont {Langer}, \citenamefont {Bollinger},
  \citenamefont {Leibfried},\ and\ \citenamefont
  {Wineland}}]{jost_entangled_2009}%
  \BibitemOpen
  \bibfield  {author} {\bibinfo {author} {\bibfnamefont {J.~D.}\ \bibnamefont
  {Jost}}, \bibinfo {author} {\bibfnamefont {J.~P.}\ \bibnamefont {Home}},
  \bibinfo {author} {\bibfnamefont {J.~M.}\ \bibnamefont {Amini}}, \bibinfo
  {author} {\bibfnamefont {D.}~\bibnamefont {Hanneke}}, \bibinfo {author}
  {\bibfnamefont {R.}~\bibnamefont {Ozeri}}, \bibinfo {author} {\bibfnamefont
  {C.}~\bibnamefont {Langer}}, \bibinfo {author} {\bibfnamefont {J.~J.}\
  \bibnamefont {Bollinger}}, \bibinfo {author} {\bibfnamefont {D.}~\bibnamefont
  {Leibfried}}, \ and\ \bibinfo {author} {\bibfnamefont {D.~J.}\ \bibnamefont
  {Wineland}},\ }\href {\doibase 10.1038/nature08006} {\bibfield  {journal}
  {\bibinfo  {journal} {Nature}\ }\textbf {\bibinfo {volume} {459}},\ \bibinfo
  {pages} {683} (\bibinfo {year} {2009})}\BibitemShut {NoStop}%
\bibitem [{\citenamefont {Ibaraki}\ \emph {et~al.}(2011)\citenamefont
  {Ibaraki}, \citenamefont {Tanaka},\ and\ \citenamefont
  {Urabe}}]{ibaraki2011}%
  \BibitemOpen
  \bibfield  {author} {\bibinfo {author} {\bibfnamefont {Y.}~\bibnamefont
  {Ibaraki}}, \bibinfo {author} {\bibfnamefont {U.}~\bibnamefont {Tanaka}}, \
  and\ \bibinfo {author} {\bibfnamefont {S.}~\bibnamefont {Urabe}},\ }\href
  {https://link.springer.com/article/10.1007/s00340-011-4463-x} {\bibfield
  {journal} {\bibinfo  {journal} {Applied Physics B}\ }\textbf {\bibinfo
  {volume} {105}},\ \bibinfo {pages} {219} (\bibinfo {year}
  {2011})}\BibitemShut {NoStop}%
\bibitem [{\citenamefont {Mordini}\ \emph {et~al.}(2023)\citenamefont
  {Mordini}, \citenamefont {Lancellotti}, \citenamefont {Negnevitsky},
  \citenamefont {Marinelli}, \citenamefont {Oswald},\ and\ \citenamefont
  {Saegesser}}]{Mordini_pytrans}%
  \BibitemOpen
  \bibfield  {author} {\bibinfo {author} {\bibfnamefont {C.}~\bibnamefont
  {Mordini}}, \bibinfo {author} {\bibfnamefont {F.}~\bibnamefont
  {Lancellotti}}, \bibinfo {author} {\bibfnamefont {V.}~\bibnamefont
  {Negnevitsky}}, \bibinfo {author} {\bibfnamefont {M.}~\bibnamefont
  {Marinelli}}, \bibinfo {author} {\bibfnamefont {R.}~\bibnamefont {Oswald}}, \
  and\ \bibinfo {author} {\bibfnamefont {T.}~\bibnamefont {Saegesser}},\ }\href
  {\doibase 10.5281/zenodo.10204606} {\enquote {\bibinfo {title} {{pytrans}},}\
  } (\bibinfo {year} {2023}),\ \bibinfo {note}
  {\url{https://doi.org/10.5281/zenodo.10204606}}\BibitemShut {NoStop}%
\end{thebibliography}


%


\iftoggle{arXiv}{
	\clearpage 
	\title{Supplemental Material: Low-excitation transport and separation of high-mass-ratio mixed-species ion chains}
    \maketitle
}{}
\label{supplement}

\section{\label{sec:trap} Segmented three-dimensional ion trap.} 
The ion trap is composed of a multi-layer three-dimensional chip made from gold-plated alumina wafers \cite{kienzler2015, deClerq2016}. The two RF electrodes are operated at \SI{\sim380}{V} with a frequency of \SI{113.5}{\mega\hertz}. The DC electrodes are controlled by a voltage range of [-10, +10] \si{\volt} and their sizes vary from \SI{155}{\micro\meter} to \SI{1000}{\micro\meter} depending on the trap region. The DC electrodes are filtered with low-pass filters with a cut-off frequency of \SI{68}{\kilo\hertz}. The ion-electrode distance is \SI{184}{\micro\meter}.

\section{\label{sec:sideband_fit} Fitting of blue sideband flops for phonon population extraction.} 
To extract the phonon population of a mode of interest we drive a blue sideband pulse of varying duration for a fixed initial internal state and subsequently read out the spin. We then fit the experimental data with a model of the form \cite{kienzler2015}
\begin{equation}
P(\downarrow,t)=\frac{1}{2}\sum_{n=0}^{n_{\text{max}}}p(n)\Big(1+\exp(-\gamma_n t)\cos(\Omega_{n,n+1}t)\Big)
\label{eq:oscillationmodel}
\end{equation}
where  $p(n)$ is the probability to find the oscillator in the $\text{n}^{\text{th}}$ energy eigenstate, $\Omega_{n,n+1}$ is the Rabi frequency for the transition between the states $\ket{\downarrow}\ket{n}$ and $\ket{\uparrow}\ket{n+1}$ and $\gamma_n$ is a phenomenological decay rate which accounts for decoherence and intensity fluctuations in the applied laser pulses. When the distribution of the probabilities $p(n)$ is thermal we use
\begin{equation}
p(n)=\frac{\bar{n}_{\mathrm{th}}^n}{\left(\bar{n}_{\mathrm{th}}+1\right)^{n+1}},
\label{eq:pn_thermal}
\end{equation}
while when $p(n)$ is a displaced thermal state distribution we use \cite{ruster2014}
\begin{equation}
\begin{aligned}
p(n)=\sum_{m=0}^{n_{max}} & \frac{\bar{n}_{\mathrm{th}}^{m}}{\left(\bar{n}_{\mathrm{th}}+1\right)^{m+1}}  e^{-|\alpha|^2}|\alpha|^{2(n+m)} m ! n ! \\
& \times\left|\sum_{l=0}^m(-1)^l \frac{|\alpha|^{-2 l}}{l !(m-l) !(n-l) !}\right|^2.
\end{aligned}
\label{eq:pn_displaced_thermal}
\end{equation}
We employ the definition 
\begin{equation}
\Omega_{n,n+1}=\Omega e^{-\eta^2/2}\sqrt{\frac{1}{n+1}}\eta L_n^1(\eta^2),
\end{equation}
where $\eta$ is the Lamb-Dicke parameter, which in our experiment is $\eta_{Be}\simeq0.4$ and $\eta_{Ca}\simeq0.06$ for the \beryllium and \calcium ions in each individual well respectively. For the Be--Ca crystal, the Lamb-Dicke parameters are $\eta_{Be}\simeq0.3$ for the mode Z2 and $\eta_{Ca}\simeq0.05$ for the mode Z1. $L_n^1(\eta^2)$ denotes the generalized Laguerre polynomial \cite{kienzler2015}. For fitting the data, we use the thermal mean phonon number $\bar{n}_{\mathrm{th}}$, the coherent mean phonon number $|\alpha|^{2}$, the decay rate $\gamma_n$ and the Rabi frequency $\Omega$ as free fit parameters, and truncate the sum in Eq.\ref{eq:oscillationmodel} to $n_{\text{max}}=20$, significantly more than the observed average phonon numbers.

Fig.\ref{fig:figS1} shows an example of data taken in the transport and separation experiments. The Be--Ca crystal is transported and subsequently separated as described in the main text. Then, a blue sideband flop is performed to extract the mean phonon number in the axial direction of the \calcium ion. Subsequently, the ions are transported in parallel to transfer the \beryllium ion into the detection zone before a readout pulse allows us to extract the excitation in the axial trap direction. The data is fitted with the model from equations \ref{eq:oscillationmodel} and \ref{eq:pn_thermal}.
\renewcommand{\thefigure}{S1}
\begin{figure}[htb]
    \centering
    \includegraphics[width=8.5cm]{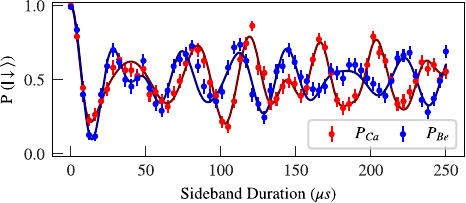}
    \caption{Extrapolation of phonon population in the axial motional mode of \calcium (red) and \beryllium (blue) after the transport and separation sequence T2--S2--PT. From the fit of eq. \ref{eq:oscillationmodel} (solid lines), we extract mean phonon numbers of $\bar{n}=\SI{1.40\pm0.08}{}$ for the calcium ion and $\bar{n}=\SI{1.44\pm0.09}{}$ for the beryllium ion.}
    \label{fig:figS1}
\end{figure}

\section{\label{sec:micromotion_compensation} Micromotion compensation} 

{\label{sec:micromotion_compensation} }
Radial micromotion is detected through the excitation of radial modes by resonant parametric modulation of the RF voltage. When the ion is far from the RF null, the motion is excited, which reduces the ion fluorescence in a subsequent detection \cite{ibaraki2011}. We compensate stray fields by applying voltages to electrodes on additional wafers of the trap stack \cite{negnevitsky2018}. To perform the compensation at any arbitrary point in the waveform sequence, the latter is run up to that position and subsequently the RF parametric modulation is switched on while setting specific compensating voltages. The sequence is then reversed and the ion fluorescence is detected. We choose the compensating voltages that prevent ion excitation. This procedure can be applied to any point in the trap, even for those without laser beam access. Before running the transport waveform, the micromotion is compensated at $11$ discrete points in the waveform which are evenly distributed over the entire transport time. During the separation waveform, the micromotion is compensated at $12$ discrete points. In between these discrete points, we linearly interpolate the micromotion compensation voltages. This calibration is performed automatically and needs to be repeated about once per month.

\section{\label{sec:applied_voltages} Voltage waveforms} 
{\label{sec:applied_voltages} }
The waveforms we employ for the transport and separation of the ion chains are generated through a numerical quadratic programming algorithm \cite{negnevitsky2018}. Once smooth ion trajectories and mode frequency profiles are defined, the solver finds the optimal voltages that minimize a cost function and satisfy specific constraints, e.g. DAC voltage limits, voltage slew rate, direction of vibrational modes relevant for cooling, etc. Additionally, the waveforms are tested and optimized experimentally. An example of ion trajectory and mode frequencies related to the transport and separation waveforms T2-S2 is shown in Fig.\ref{fig:figS2}. The Python library we developed for the generation and analysis of the waveforms has been released as an open-source package \cite{Mordini_pytrans}.

\renewcommand{\thefigure}{S2}
\begin{figure}[h]
\centering
\includegraphics[width=1\linewidth]{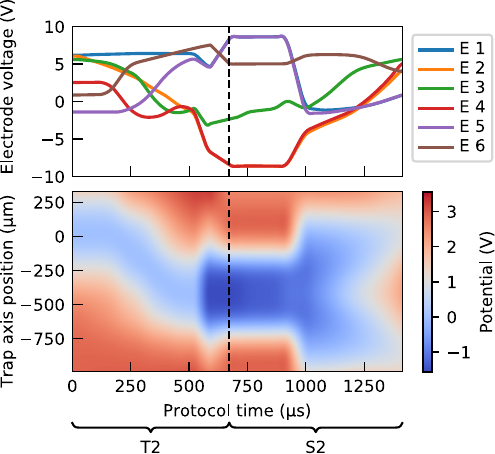} 
\caption{\label{fig:figS2} Voltages applied to the relevant trap electrodes shown in Fig.\ref{fig:fig1} (top) and generated potentials at the trap axis (bottom) while running the transport (T2) and separation (S2) waveforms. The origin of the trap position axis corresponds to the center of electrode 5, shown in the trap schematic in Fig.\ref{fig:fig1}.}
\end{figure}

\section{\label{sec:single_species_splitting} Single species separation} 
In Fig.\ref{fig:fig4}(a) of the main text we analyze the performance of the mixed-species ion chain separation while executing it at different speeds. Here we show results of similar measurements of the excitations induced by separating single-species ion crystals composed of Ca--Ca and Be--Be. The results are shown in Fig.\ref{fig:fig_sep_single_species}. In this measurement, the transport waveform works at $26\%$ lower axial frequency for a single \calcium than for the waveforms used in the main text. During this protocol, the radial mode Y1 crosses twice with the axial mode Z2 for Ca--Ca crystal without any quadrupole being applied, so the amount of coupling between the modes has not been quantified nor controlled. This crossing does not happen for Be--Be chains.
\renewcommand{\thefigure}{S3}
\begin{figure}[h]
\centering
\includegraphics[width=0.75\linewidth]{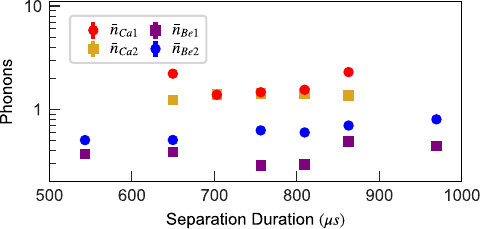} 
\caption{\label{fig:fig_sep_single_species} Separation of Ca-Ca and Be-Be crystals. The index $1$ indicates the ion trapped in the first well that reaches the detection zone after separation, while the index $2$ indicates the ion that arrives after parallel transport.}
\end{figure}

\section{\label{sec:splitting_in_other_order} Comparison of separation excitation for Be--Ca and Ca--Be} 

In the main text, we report that separating Ca--Be crystals induces higher excitations in the Z1 mode than with Be--Ca. In particular, we find excitations after the separation sequence T2-S2-PT of $\bar{n}_{Ca} {=} \SI{3.8\pm0.6}{}$ and $\bar{n}_{Be} {=} \SI{0.24\pm0.03}{}$.

We observe a few differences between the two configurations.
First, we find that the optimal value of the applied axial field depends on the crystal order, which we attribute to the presence of a pseudopotential gradient along the trap axis \cite{home2013}. Measurements of the trap frequencies of the mixed-species chains close to the critical point are shown in the upper panel of Fig.\ref{fig:figS4}. We observe that the minimum value of the Z1 frequency and the time in the waveform at which this is achieved is different in the two cases. We suspect that these differences are due to asymmetric anharmonic terms in the potential, or to the differences in position of the ions in the potential in the two different cases.
Measurements of the heating rates of two configurations at different points in the waveform are shown in the lower panel of Fig.\ref{fig:figS4}. These show a considerable difference between the two orders, which occurs at a point where the mode with the higher heating rate has a higher frequency. We do not know why this occurs, but it may relate to uncontrolled orientation of the ion chain close to the critical point. We note that radial modes (such as Y2) aligned between our DC electrodes exhibit high heating rates in the presence of a radial field and it is possible that the Ca--Be axial modes align with this at some point. 
\renewcommand{\thefigure}{S4}
\begin{figure}[htbp]
\centering
\includegraphics[width=0.75\linewidth]{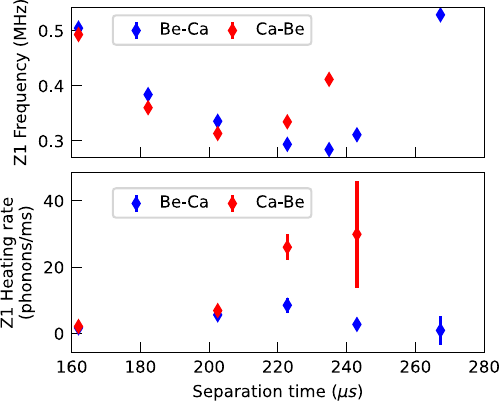} 
\caption{\label{fig:figS4} Z1 frequency and heating rate measurements for the Be--Ca and Ca--Be configurations as a function of time during the separation protocol.}
\end{figure}

\section{\label{sec:splitting_crystal} Separating larger ion crystals}
\renewcommand{\thefigure}{S5}
We also tested the separation of larger ion crystals using the same separation waveform used for the two-ion crystal employed so far. To this end, we separated an ion crystal constituted by three ions (Ca--Be--Ca and Be--Ca--Be) and four ions (Ca--Be--Ca--Be). In these measurements, we did not measure the excitation of the ions after the separation process but we did verify that the crystal can be separated arbitrarily into two parts with low loss of fluorescence. As shown in Fig.\ref{fig:figS5}, depending on an axial field applied to the ions in the separation zone, different numbers of ions can be separated to the left or right well. The results show that in this process the ion order is reliably maintained. We observe a loss of fluorescence counts only for the cases where three or more ions are in well 2.

\begin{figure}[t!]
\centering
\includegraphics[width=0.75\linewidth]{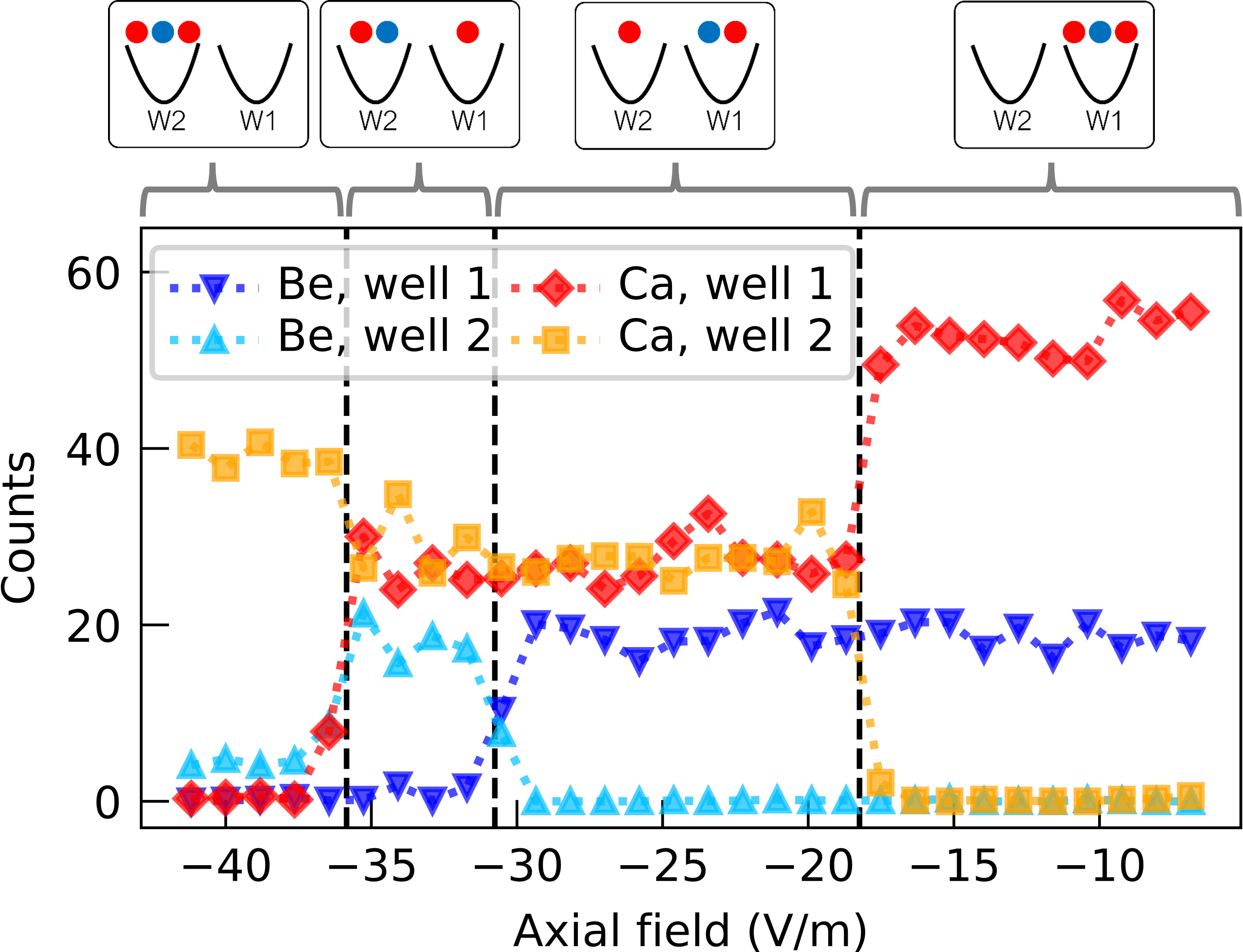} 

\vspace{0.3cm}

\includegraphics[width=0.75\linewidth]{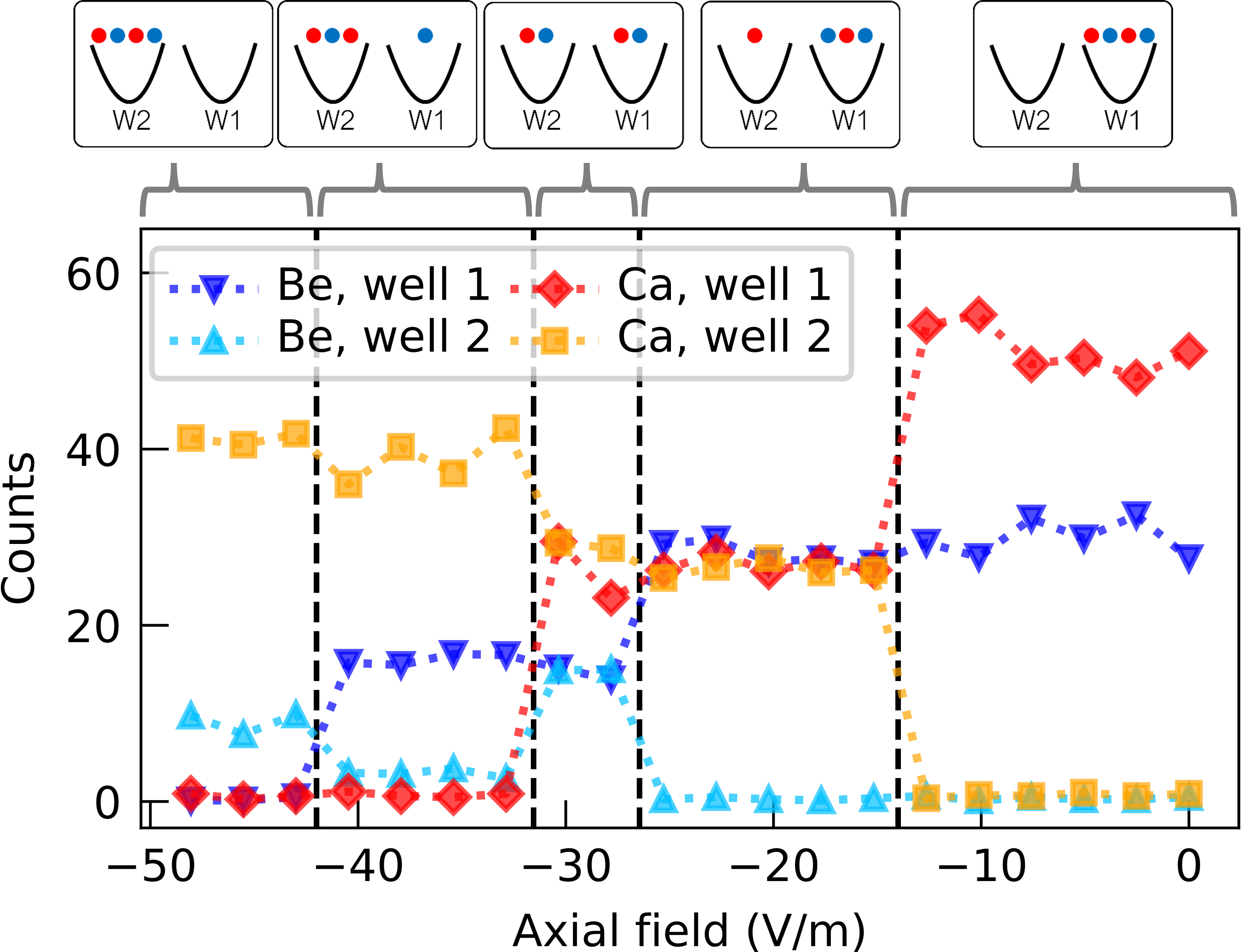} 
\caption{\label{fig:figS5} Fluorescence counts detected in the two final wells after separating Ca--Be--Ca (top) and Ca--Be--Ca--Be (bottom) ion crystals as a function of the axial field. On top of the plots, the small graphs show the final ion configurations. }
\end{figure}
\clearpage

\end{document}